\documentclass{article}

\usepackage{arxiv}

\usepackage[utf8]{inputenc}
\usepackage[T1]{fontenc}
\usepackage{hyperref}
\usepackage{url}
\usepackage{microtype}

\usepackage{graphicx}
\graphicspath{{./images/}}

\usepackage{amsmath,amssymb,amsfonts,mathrsfs,bm,stmaryrd}
\usepackage{nicefrac}
\usepackage{makecell}

\usepackage{subcaption}
\usepackage{rotating}
\usepackage{booktabs}
\usepackage{multirow}
\usepackage{algorithm}
\usepackage{algpseudocode}
\usepackage{enumitem}
\usepackage{float}
\usepackage{siunitx}
\usepackage{comment}
\usepackage{layout}
\usepackage[round,authoryear]{natbib}

\title{Bayesian Tensor-on-Tensor Varying Coefficient Model for Forecasting Alzheimer's Disease Progression}

\author{
 Yajie Liu \\
  School of Public Health \\
  The University of Texas Health Science Center\\
  Houston, TX 77030 \\
  \texttt{yajie.liu@uth.tmc.edu} \\
   \And
 Hengrui Luo \\
  Department of Statistics\\
  Rice University\\
  Houston, TX 77251 \\
  \texttt{hl180@rice.edu} \\
  \And
 Suprateek Kundu \\
  Department of Biostatistics\\
  The University of Texas MD Anderson Cancer Center\\
  Houston, TX 77030 \\
  \texttt{SKundu2@mdanderson.org} \\
  \AND
   for the Alzheimer's Disease Neuroimaging Initiative \\
}

\begin{document}
\maketitle
\begin{abstract}
We propose a novel tensor-on-tensor modeling framework that flexibly models nonlinear voxel-level relationships using Gaussian process (GP) priors, while incorporating the spatial structure of the output tensor through low-rank tensor-based coefficients. Spatial heterogeneity is captured through patch-to-voxel mappings, enabling each output voxel to depend on its spatial neighborhood. The proposed interpretable and flexible Bayesian tensor-on-tensor framework is able to capture nonlinearity, spatial information, and spatial heterogeneity. We develop an efficient Markov chain Monte Carlo (MCMC) algorithm that exploits parallel structure to sample voxel-specific GP atoms and update low-rank tensor coefficients. Extensive simulations reveal advantages of the proposed approach over existing methods in terms of coefficient estimation, inference, prediction, and scalability to high-dimensional images. Applied to longitudinal image prediction with T1-weighted MRIs from the Alzheimer's Disease Neuroimaging Initiative (ADNI), the proposed method can accurately forecast future cortical thickness. The predicted images also enable reliable prediction of brain aging, underscoring their biological relevance. Overall, the ADNI analysis highlights the model’s ability to forecast future neurobiological changes that have important implications for early detection of AD.
\end{abstract}

\keywords{
 Alzheimer's Disease \and Bayesian tensor modeling \and brain image forecasting \and Gaussian processes \and longitudinal neuroimaging}

\section{Introduction}
\label{s:intro}
Alzheimer's disease (AD), the most common neurodegenerative disease, is thought to start decades before diagnosis \citep{thompson2011design, dubois2016preclinical, knopman2021alzheimer, xiong2019harmonized}.
Since its discovery, many processes, mechanisms and biomarkers have been established \citep{knopman2021alzheimer}. However, despite
recent regulatory approvals of new AD and Related Dementia (ADRD) treatments, there is still a clear need to establish new and better treatments that, beyond affecting symptoms, can alter or reverse disease course \citep{conti2023advances}. It is of great interest to track AD progression or identify early neurobiological mechanisms predictive of future AD progression. In this context, 
Disease progression modeling (DPM) has become an important framework for characterizing biomarker dynamics and mapping the natural trajectory of diseases to predict progression \citep{koval2021ad, archetti2019multi, lorenzi2019probabilistic, maheux2023forecasting}. DPM seeks to map the natural course of neurodegenerative diseases by characterizing biomarker dynamics over time and predicting disease progression from each patient's historical data. However, DPM approaches to forecast future neurobiological changes at the voxel-level using high-dimensional imaging data at past visits,  remain limited. 


Our focus is to develop a novel Bayesian image-on-image regression (IIR) modeling approaches for forecasting brain images at future visits using images at past visits, based on longitudinal neuroimaging datasets. Subsequently, we use these forecasted images at future timepoints for predicting individuals who are most likely to experience accelerated brain aging and neurodegeneration in the future, which has direct implications in terms of early detection. While there is an increasing literature on IIR models, there are several unmet needs. For example, existing IIR approaches have been mainly designed to regress images from one modality images onto another modality, and motivated by recent advances in multimodal imaging studies. However, applications of IIR methods to longitudinal image forecasting using imaging data collected over multiple visits are limited. 

In general, IIR is a challenging problem given the high-dimensionality and complex spatial dependencies of the imaging voxels in both input and output images, resulting in difficulties when specifying model parameters. In particular, it is not straightforward to define flexible IIR models that preserve model parsimony and allow for flexible dependencies between the input and output images, while also accounting for spatially varying voxels and local spatial heterogeneity. There is a limited literature on tensor-on-tensor (TOT) modeling that provide an elegant solution to tackle IIR problems by expressing the regression coefficients as multi-dimensional tensors. TOT models allow for model parsimony via low-rank decompositions 
\citep{carroll1970analysis, tucker1966some}, while preserving spatial coherence in the coefficients connecting the input and output imaging data.

One of the earliest TOT modeling approaches appears to be proposed by Lock~\citep{TOT} that was later generalized to incorporate multiple tensor covariates \citep{gahrooei2021multiple}. Subsequent extensions were developed to incorporate tensor-train representations involving additional decompositions \citep{liu2020low}. A Bayesian TOT regression model using Tucker decomposition was proposed by ~\citep{wang2024bayesian}, with the ability to automatically infer the dimension of the core tensor. The TOT framework was extended to accommodate additional low rank decompositions and separable covariance structures to achieve further dimension reduction in \citep{llosa2022reduced}. More recently, ~\citep{varycoef} proposed a TOT varying coefficients model where both the response and covariates could be time-series of tensors. Some recent work has focused on image-on-image regressions that rely on alternative basis representations instead of leveraging tensor-based modeling frameworks. A deep learning approach was proposed in \citep{IIR} that uses a multi-stage modeling involving basis expansions in the first stage, followed by a feed-forward deep neural network to model the regression relationships between the basis coefficients in the second stage. In another paper \citep{sblf}, a sparse latent factor model was used that uses multi-level Bayesian framework involving basis representations for the outcome image followed by latent factor representations of the basis coefficients that are made to depend on the predictor images. Unlike tensor-based methods, these approaches potentially suffer from the curse of dimensionality and may not be scalable to high-dimensional 3D images. 

In spite of recent developments, there are several important gaps in TOT modeling literature. Most methods assume linear dependencies between the input and output tensors that may not be supported in practical imaging experiments. For example, images arising from different modalities may be related via complex dependencies, while longitudinal imaging experiments may involve non-linear time trends of the image evolution. For such scenarios, it is appealing to leverage  Bayesian semi-parametric regression modeling frameworks such as Gaussian process (GP) regression \citep{rasmussen2006gaussian} that have been successfully applied in a wide variety of problems in statistics literature. However, the adoption of GPs to tackle non-linearity in TOT regression models is limited, to our knowledge. 

Another critical restriction is the assumption of voxel-to-voxel mapping between the input and output images that stipulates that the variations in the output image voxel can be fully characterized by the corresponding input image voxel. However, such an assumption may fails to capture information from neighboring input voxels that impact the output voxel, and therefore fails to capture the heterogeneity due to local spatial variations \citep{lian2025patch}. A potential solution to this limitation may be found in a {\it patch-to-voxel} modeling approach that can harness the local contextual information inherent in an input patch to inform the properties of individual voxels in the output image. This type of approach has shown strong utility in diffusion MRI denoising, where multiple 3D image patches are aggregated to predict a voxel value by leveraging local spatial information \citep{fadnavis2020patch2self}. Similarly, patch-based CNN models have been applied to deformable image registration  \citep{ronneberger2015u}.  However, to our knowledge, the patch-to-voxel concept has not yet been incorporated into TOT regression frameworks. While classical TOT models could, in principle, include neighboring voxels as additional covariates, such an approach inflates the number of coefficients with neighborhood size, introduces collinearity, and retains restrictive linearity assumptions. Thus, developing a nonlinear TOT regression framework that integrates patch-to-voxel mapping to capture local spatial heterogeneity is highly desirable.

This paper addresses key gaps in TOT modeling literature by introducing a Bayesian TOT varying coefficient regression framework based on a patch-to-voxel mapping, where each voxel in the output image is associated with a surrounding 3D patch in the input image. The model employs a low-rank PARAFAC decomposition to capture global spatial patterns in tensor coefficients, combined with a multiplicative local effect that models voxel-wise non-linear dependencies between input and output images. Local non-linear effects are characterized using flexible GP priors defined independently for each voxel, allowing spatial heterogeneity through localized patch-to-voxel mappings. The proposed Bayesian framework is applicable to general input–output tensors and is further extended to sparse tensors with subject-specific sparsity patterns, motivated by neuroimaging applications.

We develop an efficient Markov chain Monte Carlo (MCMC) algorithm to implement the proposed approach that scales well to high-dimensional images containing tens of thousands of voxels per image.
The computational scalability stems from the low rank PARAFAC decomposition of the tensor coefficients, coupled with an embarrassingly parallel computation update step for the voxel-wise non-linear terms that are modeled under GP priors. The numerical advantages of the proposed approach over existing TOT approaches is illustrated via rigorous simulation studies based on 3-D images.
Our motivating applications involve Alzheimer's disease (AD) progression modeling using longitudinal Alzheimer's Disease Neuroimaging (ADNI) dataset \citep{weiner2017recent}. We use T1w-MRI scans at baseline and month 6 to train the TOT model, and subsequently perform out-of-sample prediction for images at month 12 given the month 6 images. Additionally, we also perform out-of-sample prediction for month 12 images, using month 6 images as inputs. In both types of analysis, the proposed approach shows considerable improvements in predicting voxel-wise cortical thickness (CT) at month 12 using images at previous visits, over existing approaches. Further, we demonstrate that it is possible to accurately predict accelerated brain aging at month 12 based on the corresponding predicted images for individuals with AD and mild cognitive impairment (MCI). This illustrates the ability of the proposed approach to accurately forecast future neurobiological changes given past imaging data that can be potentially prognostic of future AD progression, with important implications for AD clinical trials \citep{kim2022alzheimer}.\\

\section{Methods}
\label{s:model}

\subsection{Notations}
\label{sec:Notations}

\begin{figure}[H]
\centering
\subfloat[]{\includegraphics[width=0.8\textwidth]{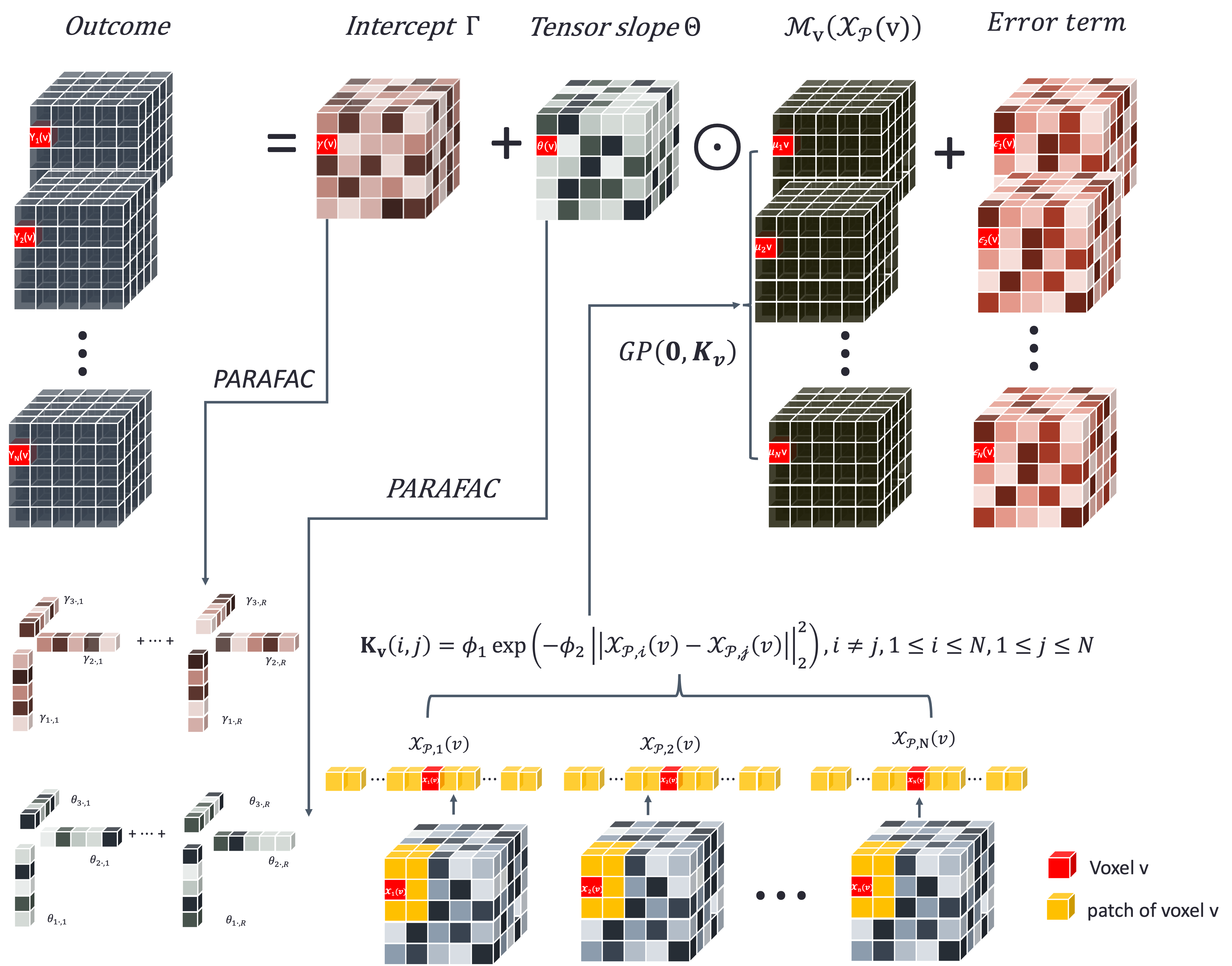}\label{fig:schematic}}\\[6pt]
\subfloat[]{\includegraphics[width=0.85\textwidth]{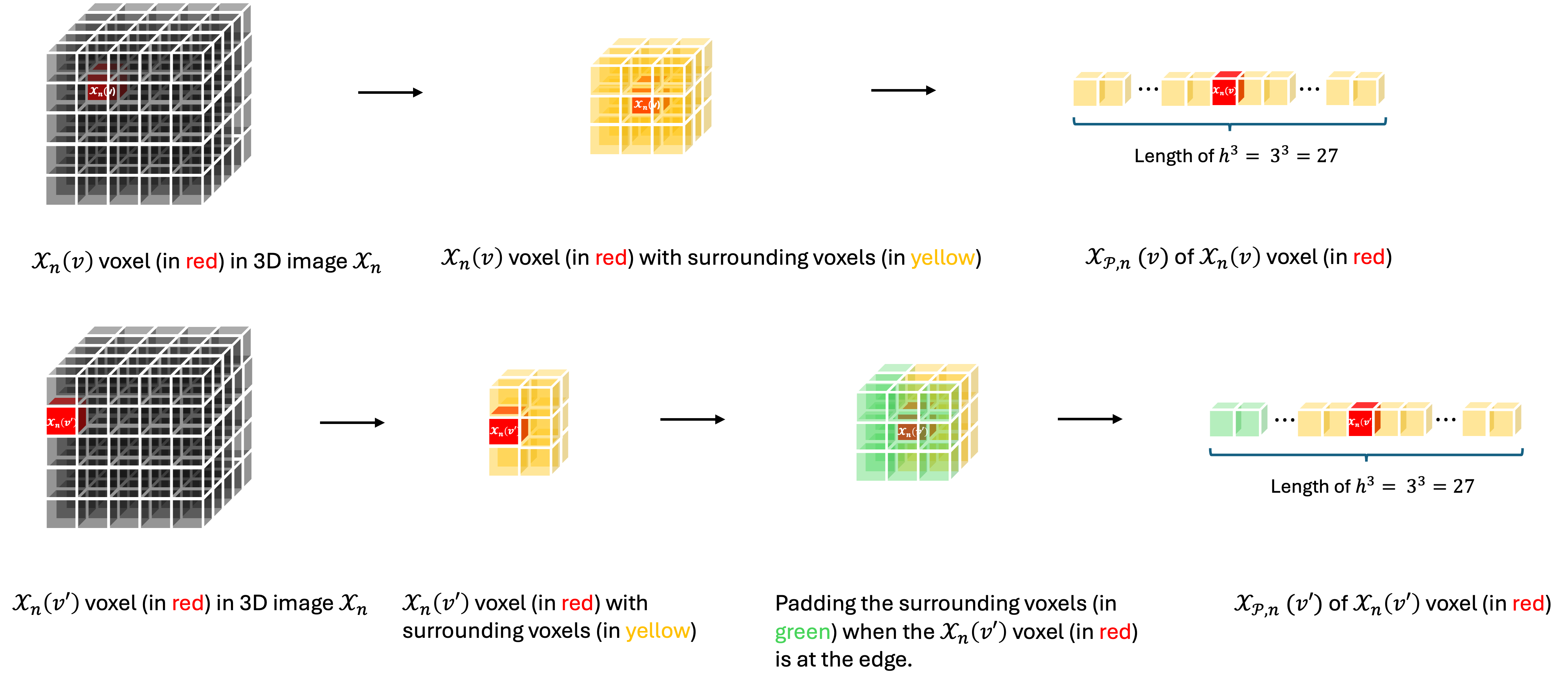}\label{fig:patch_illustrution}}
\caption{\textbf{Illustration of the BTOT-VC model structure and 3D voxel-centered patch construction.} (a) A schematic representation of the BTOT-VC model (shown without covariates for clarity). Each cube denotes a 3D tensor, with unit cubes corresponding to voxels. The model is expressed as $\mathcal{Y}_n = \Gamma + \Theta \odot \mathcal{M}_{n,\cdot}(\mathcal{X}_{\mathcal{P},n}) + \mathcal{E}_n,$ $ n=1,\ldots,N$, highlighting how voxel-wise dependencies between input and output tensors are captured through a spatially structured Gaussian process, while local information is incorporated via the patch-based mapping $\mathcal{M}_{n,v}(\mathcal{X}_{\mathcal{P},n}(v))$. (b) Illustration of voxel-centered patch extraction in a $5 \times 5 \times 5$ image. The top panel shows an interior voxel $v$ (red), for which all neighboring voxels (yellow) are observed and included in the patch $\mathcal{X}_{\mathcal{P},n}(v)$. The bottom panel shows a boundary voxel $v'$, with zero-padding (green) applied at image boundaries to ensure consistent patch dimensionality across all voxel locations.}
\label{fig:model_illustration}
\end{figure}

We consider imaging (tensor) data on $N$ samples, divided into a training set with $N_{\text{train}}$ samples and a testing set with $N_{\text{test}}$ samples. The data corresponding to the $n$th sample consists of $(\mathcal{X}_n, \mathcal{Y}_n, {\bf Z}_n)$, corresponding to $D$-dimensional input ($\mathcal{X}_n$) and output ($\mathcal{Y}_n$) tensors, and covariates ${\bf Z}_n (S\times 1)$. Our focus is on the case when the tensors correspond to images that are registered to a common template and having dimension $\mathbb{R}^{p_1 \times \cdots \times p_D}$, with $D=2,3,$ corresponding to two- and three-dimensional images, respectively. The images contain data collected over a set of spatially-varying voxels in the space $\mathcal{V}$, with the corresponding voxel-specific values being denoted as $\mathcal{X}_n(v)$ and $\mathcal{Y}_n(v)$ in the input and output images, respectively ($v\in \mathcal{V}$). Further, the spatial coordinates for each voxel are common to all images in the dataset after registration. Denote the input patch that is centered on $\mathcal{X}_n(v)$ as $\mathcal{X}_{\mathcal{P},n}(v)\in \mathbb{R}^{h\times \ldots \times h}$ of size $h^D$, noting that the patch may contain zeros corresponding to voxels near the boundary of the image/tensor. The input patch $\mathcal{X}_{\mathcal{P},n}(v)$ will be used to predict the corresponding voxel $\mathcal{Y}_n(v)$ in the output image, as described below. A schematic representation of the patch for $D=3$ and $h=3$ is illustrated in Figure \ref{fig:patch_illustrution}. We adhere to the following notational conventions: script or uppercase Greek letters (e.g., $\mathcal{M}$, $\Theta$) denote tensors; bold uppercase letters (e.g., $\mathbf{B}$) denote matrices; bold lowercase letters (e.g., $\mathbf{z}$) denote vectors. The element-wise (Hadamard) product between two tensors $\mathcal{A} \in \mathbb{R}^{p_1 \times \cdots \times p_D}$ and $\mathcal{B} \in \mathbb{R}^{p_1 \times \cdots \times p_D}$ is denoted as $\mathcal{A} \odot \mathcal{B}$, and denote the Euclidean norm of a vector as $|| \cdot ||_2$.

\textbf{Tensor notations:} Tensor decompositions provide frameworks for achieving parameter parsimony and capturing spatial correlation between voxels in image data \citep{kolda2009tensor,kundu2024flexible,li2017parsimonious,li2018tucker,zhou2013tensor,luo2025efficient}.
We employ the PARAFAC decomposition, also known as the CANDECOMP/PARAFAC (CP) decomposition that factorizes a tensor into a sum of $R$ rank-one tensors \citep{kolda2009tensor}. Mathematically, the CP decomposition expresses a tensor $\mathcal{A} \in \mathbb{R}^{p_1 \times \cdots \times p_D}$ of rank $R$ as a sum of outer products of 1-dimensional tensor margins along different modes of the tensor:
$ \mathcal{A} = \sum_{r=1}^R \mathbf{a}_{1\cdot,r} \circ \mathbf{a}_{2\cdot,r} \circ \cdots \circ \mathbf{a}_{D\cdot,r}$,  where $\circ$ denotes the outer product, and $\mathbf{a}_{d\cdot,r} \in \mathbb{R}^{p_d \times 1}$ denotes the tensor margin (vector) along mode $d$ of the tensor ($d=1,\cdots,D$), and $R$ denotes the tensor rank. The PARAFAC decomposition in our model achieves parameter parsimony by massively reducing the number of parameters from $\prod_{d=1}^D p_d$ to $R(p_1 + \ldots + p_d)$. It simultaneously captures the spatial structure of the imaging data \citep{kundu2018bayesian}. While the tensor margins are themselves not identifiable, the overall tensor coefficient expressed as an outer product of tensor margins is typically identifiable \citep{guhaniyogi2021bayesian}. The $(i_1,i_2,\ldots, i_D)$th entry of the tensor $\mathcal{A}$ is denoted as $\mathcal{A}_{(i_1,i_2,\ldots, i_D)}$. When the tensors correspond to images ($D=2,3$), each tensor entry $(i_1,i_2,\ldots, i_D)$ represents a unique voxel $v \in \mathcal{V}$, where $\mathcal{V}$ denotes the space of all tensor indices $\{(i_1,i_2,\ldots, i_D): 1\le i_d \le p_d, d=1,\ldots,D\}$. Our motivating applications consider tensor representation of images, where each index $(i_1,i_2,\ldots, i_D)$ (or equivalently  $v \in \mathcal{V}$) correspond to a certain spatial coordinate that is common across all images/samples after registration.  

The vectorization operator on the tensor is defined as $vec(\mathcal{A}) $ as stacking the tensor into a long vector of dimension $V=\prod_{d=1}^D p_d$ with the elements in the vector being organized as $vec(\mathcal{A})[i_1 + \sum_{d=2}^D (\prod_{k=1}^{d-1} p_k)(i_d-1)] = \mathcal{A}_{(i_1,i_2,\ldots, i_D)}$. In another word, $V$ is the size of voxel space, denotes as $V = |\mathcal(V)|$. Define the slice of a tensor (denoted as $\mathcal{A}_{\cdot,dj}$) as a reduced tensor of dimension $\prod_{d'\ne d, d'=1}^D p_{d'}$ that is obtained by fixing the index $j\in \{1,\ldots,p_d \}$ along the $d$th mode of the tensor. The scalar product between two tensors of dimension $D$ is written as $\langle \mathcal{A}, \mathcal{B} \rangle = \sum_{i_1,i_2,\ldots, i_D}\mathcal{A}_{(i_1,i_2,\ldots, i_D)} \mathcal{B}_{(i_1,i_2,\ldots, i_D)} $. Similarly, a scalar product between two tensor slices $\mathcal{A}_{\cdot,dj}$ and $\mathcal{B}_{\cdot,dj}$ is defined as $\langle \mathcal{A}_{\cdot,dj}, \mathcal{B}_{\cdot,dj} \rangle = \sum_{i_1,\ldots, i_{d-1}, i_d=j, i_{d+1},\ldots i_D}\mathcal{A}_{(i_1,i_2,\ldots, i_D)} \mathcal{B}_{(i_1,i_2,\ldots, i_D)} $ that holds mode $d$ fixed at the $j$th element $(j\in \{1,\ldots,p_d \})$, while taking the product over all remaining dimensions. 

\subsection{Proposed Tensor-on-Tensor Varying Coefficients Regression Model}
We propose the following semi-parametric Bayesian approach that relaxes commonly used linearity assumptions in TOT models by introducing a non-linear varying coefficient term. We write the model as:
\begin{equation}
    \mathcal{Y}_n = \Gamma + \Theta \odot \mathcal{M}_{n,\cdot}(\mathcal{X}_{\mathcal{P},n}) + \sum_{s=1}^S \mathcal{D}_s  z_{ns} + \mathcal{E}_n, \quad n = 1, \dots, N,
    \label{eq:tensor}
\end{equation}
where $\Gamma \in \mathbb{R}^{p_1 \times \cdots \times p_D}$ is the intercept tensor that captures common structural information across all images, $\Theta \in \mathbb{R}^{p_1 \times \cdots \times p_D}$ is the slope tensor that captures the spatial patterns in the response tensor, $\mathcal{M}_{n,\cdot}(\mathcal{X}_{\mathcal{P},n}) \in \mathbb{R}^{p_1 \times \cdots \times p_D}$  denotes the non-linear tensor varying coefficient term that encodes the subject-specific patch-to-voxel dependencies for modeling the unknown relationships between output tensor and input patch $\mathcal{X}_{\mathcal{P}}$, $\mathcal{D}_s \in \mathbb{R}^{p_1 \times \cdots \times p_D}$ corresponds to the tensor regression coefficients to capture the covariate effects, and $\mathcal{E}$ denotes the residual term that follows a Gaussian distribution with $vec(\mathcal{E}_n) \stackrel{i.i.d.}{\sim} N_{V}(0,\sigma^2_e I_{V})$, where $N_V({\bm\mu},\Sigma)$ refers to a $V$-dimensional Gaussian distribution with mean ${\bm\mu}(V\times 1)$ and covariance $\Sigma(V\times V)$, and $z_{ns}$ denotes the $s$-th element of $Z_n$ for the $n$-th subject. The proposed model (\ref{eq:tensor}) operates by capturing global structural information (spatial information for images) in the tensor intercept and slope ($\Gamma,\Theta$) via a low-rank PARAFAC decomposition (described in sequel), while simultaneously encoding the local non-linear voxel-level dependencies between the input and output images, independently across tensor units/voxels via the $\mathcal{M}(\cdot)$ parameters. The PARAFAC decomposition for tensor coefficients is: 
\begin{align}    
\Gamma = \sum_{r=1}^R \boldsymbol{\gamma}_{1\cdot,r} \circ \cdots \circ \boldsymbol{\gamma}_{D\cdot,r}, \mbox{ } \Theta = \sum_{r=1}^R \boldsymbol{\theta}_{1\cdot,r} \circ \cdots \circ \boldsymbol{\theta}_{D\cdot,r}, \mbox{ } \mathcal{D}_s = \sum_{r=1}^{R} \boldsymbol{\delta}_{1\cdot,r,s} \circ \cdots \circ \boldsymbol{\delta}_{D\cdot,r,s}, 
\label{eq:gamma_margin}
\end{align}

where the tensor rank $R$ is assumed the same across $\Gamma$ and $\Theta$ for simplicity, but can be varied  \citep{kundu2023bayesian}.  
The tensor margins are identifiable only up to a permutation and a multiplicative constant, unless additional constraints are imposed \citep{kundu2023bayesian}. 
Moreover, the tensor slopes $\Theta$ are also not identifiable in model (\ref{eq:tensor}). However, the product $\Theta \odot \mathcal{M}_{n,\cdot}$ is identifiable, and this will be the parameter of interest for modeling purposes.

Model (\ref{eq:tensor})  can be rewritten in terms of a simpler voxel level representation as follows, where each $v \in \mathcal{V}$ corresponds to a certain tensor index $(i_1,\ldots,i_D)$:
\begin{equation}
    \mathcal{Y}_{n}(v) = \Gamma(v) + \Theta(v) \mathcal{M}_{n,v}(\mathcal{X}_{\mathcal{P},n}(v)) + \sum_{s=1}^S \mathcal{D}_s(v)  z_{ns} + \mathcal{E}_n(v), \quad n = 1, \dots, N,\; v \in \mathcal{V},
    \label{eq:voxel-wise}
\end{equation}
where $\mathcal{M}_{n,v}(\mathcal{X}_{\mathcal{P},n}(v)):\mathbb{R}^{h^D}\to \mathbb{R}$ corresponds to a non-linear term that maps the input tensor patch $\mathcal{X}_{\mathcal{P}}(v)$ of size $h^D$ to the scalar output tensor unit/voxel $\mathcal{Y}_{n}(v)$. This term is subject- and voxel-specific, enabling us to incorporate spatial-heterogeneity embedded in the patch-to-voxel mappings. It is modeled using a Gaussian process prior  as follows: 
\begin{eqnarray}
 \mathcal{M}_{\cdot,v}(\mathcal{X}_{\mathcal{P}}(v)) \stackrel{indep}{\sim} GP(\bm{0}, \mathbf{K}_{v}),  \mbox{ }  \mathbf{K}_{v}(i,j) = \phi_{1} \exp\left(-\phi_{2} || \mathcal{X}_{\mathcal{P},i}(v) - \mathcal{X}_{\mathcal{P},j}(v)||_2 ^2 \right),   \mbox{ } v \in \mathcal{V}, \label{eq:voxel-GP}
\end{eqnarray}
where the GP is defined independently for each voxel, with the corresponding squared exponential covariance kernel given as $\mathbf{K}_{v}(N\times N)$ with $\phi_1\sim \mbox{Inv-Ga}(a_{\phi_1}, b_{\phi_1})$  corresponding to the variance/scale parameter and $\phi_2>0$ denoting the lengthscale parameter that controls the smoothness of the GP and is assigned a non-informative prior ($\pi(\phi_2) \propto 1$).

 While the number of distinct $\mathcal{M}$ parameters increases with $N$ and the tensor size ($V$), this is the price to pay for flexibly accommodating non-linear effects with local spatial heterogeneity. However, it is possible to parallelize the computation across voxels when sampling $\mathcal{M}$, resulting in computational scalability (see sequel). Moreover, $\mathcal{M}$ is integrated out while performing out-of-sample prediction using posterior predictive distributions, as per convention. We denote the proposed model as Bayesian tensor-on-tensor varying coefficient model (BTOT-VC), and a schematic is presented in Figure \ref{fig:schematic}.

\noindent{\textbf{Prior on Tensors:}} We adopt a hierarchical Bayesian approach in tensor modeling, 
based on results in existing literature \citep{kundu2023bayesian}. To enable spatial smoothing, we assign normal priors to the tensor margins in (\ref{eq:gamma_margin}) that incorporate correlations between neighboring tensor units. 
For mode $d$ and component $r$, the priors on the tensor margins are specified:
\begin{equation} 
    \boldsymbol{\gamma}_{d\cdot,r} \sim \mathcal{N}(0, \tau^\gamma \mathbf{W}_{d,r}^\gamma), \mbox{ }
    \boldsymbol{\theta}_{d\cdot,r} \sim \mathcal{N}(0, \tau^\theta \mathbf{W}_{d,r}^\theta), \mbox{ }
    \boldsymbol{\delta}_{d\cdot,r,s} \sim \mathcal{N}(0, \tau_s^\delta \mathbf{W}_{d,r,s}^\delta), d = 1, 2, \dots, D,
    \label{eq:prior_tensor}
\end{equation}
where $r=1,\ldots,R$, $\tau^\gamma$, $\tau^\theta$, and $\tau_s^\delta$ are global scale/variance parameters that control overall shrinkage for  $\boldsymbol{\gamma}_{d\cdot,r},\boldsymbol{\theta}_{d\cdot,r},\boldsymbol{\delta}_{d\cdot,r,s}$, respectively, and are sharing the same Gamma distribution prior $\tau \sim \text{Ga}(a_\tau, b_\tau)$. The covariance matrices $\mathbf{W}_{d,r}^\gamma$, $\mathbf{W}_{d,r}^\theta$, $\mathbf{W}_{d,r,s}^\delta$ for the tensor margins $\boldsymbol{\gamma}_{d\cdot,r}$, $\boldsymbol{\theta}_{d\cdot,r}$ and $\boldsymbol{\delta}_{d\cdot,r,s}$ are $p_d \times p_d$ positive semi-definite matrices that encode spatial correlations along the $d$-th tensor mode, where $p_d$ denotes the length of the tensor margins at the $d$-th mode. To prevent excessive growth in model parameters as $D$ increases and simultaneously encourage spatial smoothing, we adopt a shared parametric correlation structure for $\mathbf{W}_{d,r}$. For example, we specify $\mathbf{W}_{d,r}^\gamma$ as:
$ \mathbf{W}_{d,r}^\gamma = w_{d,r}^\gamma \Lambda_{d,r}^\gamma,$
where $\Lambda_{d,r}^\gamma$ encodes spatial correlations such that $\Lambda_{d,r}^\gamma(k_1, k_2) = \exp\{-\alpha_{d,r}^\gamma|k_1 - k_2|\}$, indicating that prior correlations decrease as the distance between the $k_1$-th and $k_2$-th elements increases, for the $d$-th margin $(k_1, k_2 = 1, \cdots, p_d)$. The correlations are further modulated by the lengthscale parameter $\alpha_{d,r}^\gamma \sim \text{Ga}(a_\alpha, b_\alpha)$. A larger value of $\alpha_{d,r}^\gamma$ implies a weaker correlation. 
To model the diagonal variance terms, hierarchical priors are imposed as $w_{d,r}^\gamma \sim \text{Exp}(\lambda_{d,r}^\gamma / 2), \quad \lambda_{d,r}^\gamma \sim \text{Ga}(a_\lambda, b_\lambda).$ Consequently, $\mathbf{W}_{d,r}^\gamma(k_1, k_2)$ is expressed as $w_{d,r}^\gamma \exp(-\alpha_{d,r}^\gamma |k_1 - k_2|)$. The covariance matrices for the other regression coefficients in Equation (\ref{eq:prior_tensor}) are constructed in a similar manner: $ \mathbf{W}_{d,r}^\theta(k_1,k_2) = w_{d,r}^\theta\exp(-\alpha_{d,r}^\theta|k_1-k_2|)$, $ \mathbf{W}_{d,r,s}^\delta(k_1,k_2) = w_{d,r,s}^\delta\exp(-\alpha_{d,r,s}^\delta|k_1-k_2|)$, with the priors on $w_{d,r}^\theta,$ and $w_{d,r,s}^\delta$ being defined similarly as $\pi(w_{d,r}^\gamma)$. 
For simplicity, a common tensor rank $R$ is assumed for all coefficient tensors, including $\Gamma, \Theta, \mathcal{D}_s$. The tensor rank $R$ is selected using the deviance information criterion (DIC), a Bayesian model selection criterion that balances model fit and complexity by accounting for the effective number of parameters \citep{guhaniyogi2017bayesian}. The expression for DIC is provided in Section~\ref{s:dic}.

\subsection{Posterior Predictive Distributions}
The proposed approach is trained on a training set with $N_{\text{train}}$ samples, and the out-of-sample performance is evaluated on a test sample with $N_{\text{test}}$ samples. Let us denote the vector of training sample tensors at the $v$th voxel as 
$\mathcal{Y}_{\text{train},n}(v):= [ \mathcal{Y}_{1}(v), \ldots, \mathcal{Y}_{N_{\text{train}}}(v)]^\top$ and similarly denote the corresponding collection of test sample tensors as $\mathcal{Y}_{\text{test},n}(v)$. Further, denote the concatenated vector at the $v$th voxel as $\mathcal{Y}(v)= (\mathcal{Y}^\top_{\text{train},n}(v),\mathcal{Y}^\top_{\text{test},n}(v))^\top$ that is of length $N=N_{\text{train}}+N_{\text{test}}$. Integrating out the GP atoms, the joint distribution for $\mathcal{Y}(v)$ is: 
\begin{equation}
    \mathcal{Y}(v) \sim \mathcal{N}\left( \Gamma(v) + \sum_{s=1}^S \mathcal{D}_s(v)  \mathbf{Z}_{s},\ \Theta(v)^2 \mathbf{K}_{v} + \sigma_e^2 \text{diag} \begin{bmatrix} \mathbf{1}_{N_{\text{train}}} \\ \mathbf{0}_{N_{\text{test}}} \end{bmatrix} \right), v \in \mathcal{V},
    \label{eq:y_model}
\end{equation}
where the test samples are assumed noiseless \citep{wang2023intuitive}, and $\mathbf{K}_{v}$ is  written as: 
\[
\mathbf{K}_{v} \;=\;
\begin{bmatrix}
\mathbf{K}_{v,TT} & \mathbf{K}_{v,T*}\\[2pt]
\mathbf{K}_{v,*T} & \mathbf{K}_{v,**}
\end{bmatrix},
\mbox{}
\begin{aligned}
&\mathbf{K}_{v,TT}\in\mathbb{R}^{N_{\text{train}}\times N_{\text{train}}},
\mathbf{K}_{v,T*}\in\mathbb{R}^{N_{\text{train}}\times N_{\text{test}}},
\mathbf{K}_{v,**}\in\mathbb{R}^{N_{\text{test}}\times N_{\text{test}}},
\end{aligned}
\]
where $\mathbf{K}_{v,TT}$ captures the correlations between training samples,  $\mathbf{K}_{v,**}$ captures the correlations between test samples, and $\mathbf{K}_{v,T*}$ captures the correlations between training and test samples, corresponding to voxel $v$. The corresponding conditional distribution is:
\begin{equation}
\mathcal{Y}_{\text{test}}(v)\mid \mathcal{Y}_{\text{train}}(v), \mathbf{Z}\sim
\mathcal{N}\!\Big(
m_{\text{test}}(v) + \mathbf{K}_{vo}^{\!\top} \mathbf{K}_{vv}^{-1}\!\big[\mathcal{Y}_{\text{train}}(v)-m_{\text{train}}(v)\big],
\mathbf{K}_{oo} - \mathbf{K}_{vo}^{\!\top} \mathbf{K}_{vv}^{-1} \mathbf{K}_{vo}
\Big), 
\label{eq:krigging}
\end{equation}
where $
\mathbf{K}_{vv} = \Theta(v)^2\,\mathbf{K}_{v,TT}+ \sigma_e^2 \mathbf{I}_{N_{\text{train}}},\mbox{ }
\mathbf{K}_{vo} = \Theta(v)^2\,\mathbf{K}_{v,T*},\mbox{ }
\mathbf{K}_{oo} = \Theta(v)^2\,\mathbf{K}_{v,**}
$, with GP predictive mean and covariance are given as 
$
    m_{\text{train}}(v) = \Gamma(v){\bf 1}_{N_{\text{train}}} + \sum_{s=1}^S \mathcal{D}_s(v)  \mathbf{Z}_{\text{train},s}, \mbox{}
    m_{\text{test}}(v) = \Gamma(v){\bf 1}_{N_{\text{test}}} + \sum_{s=1}^S \mathcal{D}_s(v) \mathbf{Z}_{\text{test},s}
$, and ${\bf 1}_N$ denotes a vector of ones of length $N\times 1$. Prediction for the test set values is conducted using a Kriging approach that leverages the conditional distribution  $\pi(\mathcal{Y}_{\text{test}}(v)\mid \mathcal{Y}_{\text{train}}(v), \Gamma(v), \Theta(v),  \mathbf{Z}_{\text{train},s},\sigma^2_e) $ as defined in (\ref{eq:krigging}).

\subsection{Extension to Sparse Tensor-on-Tensor Modeling}
Our motivating applications using neuroimaging datasets often involve sparse images/tensors where the sparsity patterns may vary across samples. For example, we analyze voxel-wise CT features derived from T1w-MRI scans, which are extracted for spatially distributed gray matter in the brain but are absent for white matter regions or cerebrospinal fluid (CSF), resulting in sparsity. To extend our model to accommodate sparse tensors, we incorporate masks into  our  modeling framework that are directly pre-specified using the neuroimaging dataset. The brain mask specifies units/voxels that lie in relevant brain regions that proffer non-zero neuroimaging features. We assume the same brain mask for the input and output tensors for a given sample in our modeling framework, but the masks are allowed to vary across samples. We extend model (\ref{eq:tensor}) to incorporate subject-specific masks by `zeroing-out' global tensor coefficients corresponding to voxels lying in the common mask across samples, and restricting the non-linear mapping $\mathcal{M}$ to the set of voxels within the common mask, while adopting a simple linear mapping corresponding to all voxels lying outside the common mask but within subject-specific masks. This strategy allows to preserve subject-specific sparsity patterns while maintaining non-linear patch-to-voxel mapping under the proposed Bayesian TOT varying coefficient regression model.

 We define the subject-specific mask as the set of all units/voxels such that $\mathcal{S}_n = \{v: \mathcal{X}_n(v) \ne 0\}$, and define the corresponding binary tensor $\mathcal{V}_{M_n} \in \{0,1\}^{p_1\times\cdots\times p_D}$ with elements 
as: $\mathcal{V}_{M_n}(v) = I(v\in \mathcal{S}_n)$ that contains zero elements corresponding to all tensor units lying outside the mask $\mathcal{S}_n$, where $I(\cdot)$ denotes an indicator function. 
Further, we define a group-level mask as the set of all voxels that contain non-zero measurements for at least $100\tau\%$ samples in the dataset, i.e. 
 $\mathcal{S}_0:= \{v: I(\mathcal{X}_n(v) \ne 0)/N  \ge \tau\}$, and define the corresponding group-level binary tensor $\mathcal{V}_M \in \{0,1\}^{p_1\times\cdots\times p_D}$ with elements as $ \mathcal{V}_M(v) = I(v\in \mathcal{S}_0)$.
The group level mask captures consistent or global sparsity patterns across subjects by identifying voxels that lack meaningful cortical thickness information across the majority ($100\tau\%$) of subjects. The regions outside the mask correspond to uninformative or background regions that are not relevant for modeling, and may introduce noise into model estimation if included. We choose $\tau$ to be a high number (eg: $\tau=80\%$, although other percentages could be considered). By construction, the group level masks contain a subset of voxels/units included in the subject-specific masks, i.e. $\mathcal{S}_0 \subset \cup_{n=1}^N \mathcal{S}_n$, with $\cup_{n=1}^N \mathcal{S}_n \setminus \mathcal{S}_0$ denoting the voxels lying outside the group-level mask, where $A \setminus B$ indicates the set difference.

Let us denote $\mathcal{Y}_{n,\mathcal{M}_n}$ as the response tensor for the $n$th sample that is restricted to the set of units/voxels within the subject-specific mask $\mathcal{S}_n$. Using these masks, the model in Equation~\eqref{eq:tensor} is modified as:
\begin{equation}
    \mathcal{Y}_{n,\mathcal{V}_{M_n}} = \Gamma \odot \mathcal{V}_{M_n} 
    + (\Theta \odot \mathcal{V}_{M_n}) \odot \mathcal{M}_{n, \mathcal{V}_M}(\mathcal{X}_{\mathcal{P},n}) 
    + \sum_{s=1}^S (\mathcal{D}_s \odot \mathcal{V}_{M_n})z_{ns}  
    + \mathcal{E}_{n,\mathcal{V}_{M_n}},
    \label{eq:tensor_mask}
\end{equation}
where $(\mathcal{M}_{1, \mathcal{V}_M}(\mathcal{X}_{\mathcal{P},1})(v), \ldots, \mathcal{M}_{N, \mathcal{V}_M}(\mathcal{X}_{\mathcal{P},N})(v))$ represents a non-linear patch-to-voxel mapping that is modeled under a Gaussian process as in (\ref{eq:voxel-GP}) for all voxels $v\in \mathcal{S}_0$, while for all 
$v\in \cup_{n=1}^N \mathcal{S}_n \setminus \mathcal{S}_0$, we assume a simple linear mapping 
$\mathcal{M}_{n, \mathcal{V}_M}(\mathcal{X}_{\mathcal{P},n}) =\mathcal{X}_n(v)$. 
This enables us to model complex dependencies between the input and output images corresponding to voxels that are observed consistently across all samples ($\mathcal{S}_0$), while resorting to a simple linear association between the input and output voxels that represent subject specific sparsity patterns and lie outside of $\mathcal{S}_0$. Moreover, the intercept and slope tensors as well as the covariate contributions are multiplied (element-wise) with subject-level binary tensors $\mathcal{V}_{\mathcal{M}_n}$, which serves to zero-out the effect of zero voxels when estimating these parameters.  The resulting model specification in (\ref{eq:tensor_mask}) preserves the sparsity patterns in the observed data in the tensor-on-tensor modeling, uses data on all samples and observed voxels to compute global tensor coefficients ($\Gamma,\Theta,\mathcal{D}_1,\ldots,\mathcal{D}_s$). Further, it restricts the non-linear dependencies to a subset of units/voxels included in the group-level mask $\mathcal{S}_0$, while specifying simple linear mappings for fringe voxels lying outside $\mathcal{S}_0$.

\subsection{Posterior Computation} 
We estimate the unknown parameters in Equation~\eqref{eq:tensor} by sampling from their posterior distributions using MCMC. Most parameters, including the tensor margins of $\Gamma$, $\Theta$, and $\mathcal{D}_s$, as well as global scale parameters ($\tau$, $\lambda$, $\sigma^2$), and the variance in GP ($\phi_1$), have conjugate full conditional distributions and are efficiently updated via Gibbs sampling. However, certain hyperparameters do not admit closed-form posteriors and are updated using Metropolis-Hastings (MH) steps. These include the tensor margin length-scale parameters $\alpha_{d,r}^\gamma$, $\alpha_{d,r}^\theta$, $\alpha_{d,r,s}^\delta$, and the GP kernel length-scale parameter $\phi_2$. For the tensor covariance parameters $\alpha$, we employ a fixed-variance log-Normal random walk proposal of the form:
$\alpha_{d, r, s_x+1}^\gamma \mid \alpha_{d, r, s_x}^\gamma \sim \text{log-Normal}(\alpha_{d, r, s_x}^\gamma, \sigma_\alpha^2)$.
For the GP kernel parameter $\phi_2$, we implement a log-Normal random-walk MH update $\phi_{2, s_x+1} \mid \phi_{2, s_x} \sim \text{log-Normal}(\phi_{2, s_x}, \sigma_{\phi_2}^2)$, where the $s_x$ index the MCMC iteration. The proposal variance  $\sigma_{\phi_2}^2$ is dynamically tuned to achieve an optimal acceptance rate, while ensuring the validity of the Markov property by avoiding inadmissible adaptation strategies \citep{adaptive}. The GP terms $\mathcal{M}_{n,v}(\mathcal{X}_{\mathcal{P},n}(v))$ can be drawn simultaneously across all voxels using an embarrassingly parallelized structure. We performed 10{,}000 MCMC iterations, setting the first 50\% as burn-in. Convergence was assessed using Geweke's diagnostic. 
Full details of the MCMC algorithm and parameter updates are provided in the Section \ref{s:posterior_computation}. We summarize the full MCMC procedure in Algorithm~\ref{alg:MCMC}.

\begin{algorithm}
\caption{MCMC steps for BTOT-VC}
\label{alg:MCMC}
\begin{algorithmic} [1]
\State Update the tensor margins and hyperparameters.
\Statex \quad 1) Update tensor margins of $\Gamma$, $\Theta$ and $\mathcal{D}_s$ given $\mathcal{M}_{n,v}(\mathcal{X}_{\mathcal{P},n}(v))$ from the last iteration. 
\Statex \quad 2) Draw the covariance matrices $W_{d,r}$, the rate parameters $\lambda_{d,r}$, global variance scale parameter $\tau$ from posterior distributions. 
\Statex \quad 3) Draw the parameters $\alpha_{d,r}$ using a MH step.
\State Update $\mathcal{M}_{n,v}(\mathcal{X}_{\mathcal{P},n}(v))$ terms and hyperparameter in GP. 
\Statex \quad 1) Update the variance $\phi_1$ in kernel matrices $\mathbf{K}_v$ based on its posterior distribution. 
\Statex \quad 2) Update the lengthscale $\phi_2$ using an adaptive MH step. 
\Statex \quad 3) Update the kernel matrices $\mathbf{K}_v$ based on the updated variance and lengthscale parameters $\phi_1$ and $\phi_2$. (Equation \ref{eq:voxel-GP})
\State Update the variance $\sigma_e^2$ of error residuals $\mathcal{E}_n$ based on its posterior distribution given the updated $\Gamma$, $\Theta$, $\mathcal{D}_s$, $K_v$ and $\mathcal{M}_{n,v}(\mathcal{X}_{\mathcal{P},n}(v))$. 
\State Predict $\mathcal{Y}_{\text{test}}(v)$ using kriging approach. (Equation \ref{eq:krigging})
\end{algorithmic}
\end{algorithm}

\section{Simulation}
\label{s:simu}
\subsection{Data Generation} 
\subsubsection{Outcome image generation}
To evaluate the performance of the proposed approach, we conducted rigorous simulation studies using synthetic 3D ($32 \times 32 \times 32$, and $8 \times 8 \times 8$) images generated without sparsity. We did not incorporate baseline covariates when simulating the data, to make the comparisons with competing methods equitable.
We considered three distinct scenarios to characterize different types of relationships between the predictor and outcome images using model (\ref{eq:tensor}) that incorporates CP decomposition for global tensor coefficients.

{\noindent \bf Scenario 1:} Model (\ref{eq:tensor}) was used to generate   $\mathcal{Y}_n(v)$ by using an unknown mapping (modeled under a GP) on a local patch of voxels with dimension $3 \times 3 \times 3$ in $\mathcal{X}_{\mathcal{P},n}(v)$. The resulting data generating mechanism incorporates nonlinear dependencies and local spatial heterogeneity.

{\noindent \bf Scenario 2:} Model (\ref{eq:tensor}) was used to generate $\mathcal{Y}_n$ incorporating nonlinear voxel-wise dependencies such that $\mathcal{Y}_n(v)$ only depended on $\mathcal{X}_n(v)$. The data generating structure does not use input image patches, and therefore ignores local spatial heterogeneity, which resembles most existing tensor-on-tensor approaches in literature.

{\noindent \bf Scenario 3:} Model (\ref{eq:tensor}) was used to generate $\mathcal{Y}_n(v)$ under a pre-specified parametric nonlinear mapping, defined as $\mathcal{M}_{n,v}(\mathcal{X}_n(v)) = \log(1 + \mathcal{X}_n(v)^2)$. The resulting data generating mechanism incorporates nonlinear voxel-wise relationships without leveraging input image patches, and therefore does not account for local spatial heterogeneity.

{\noindent \bf Scenario 4:} Model (\ref{eq:tensor}) was used to generate $\mathcal{Y}_n(v)$ under a linear mapping, defined as $\mathcal{M}_{n,v}(\mathcal{X}_n(v)) = \mathcal{X}_n(v)$. The resulting data generating mechanism captures linear voxel-wise dependencies and does not incorporate input image patches or local spatial heterogeneity.

\noindent
\subsubsection{Input image generation} We also considered four distinct strategies for generating the input images $\mathcal{X}_n$: (a) images generated from standard normal distribution, i.e., $\mathcal{X}_n(v) \stackrel{i.i.d}{\sim} \mathcal{N}(0, 1)$; (b) images generated from an Uniform distribution, i.e. $\mathcal{X}_n(v) \stackrel{i.i.d}{\sim} \mathcal{U}(0, 1)$; (c) images generated using wavelets - a Daubechies least asymmetric wavelet with four vanishing moments and a decomposition level of $j_0 = 3$ was used, with wavelet coefficients generated from a standard normal distribution; and (d) images were generated from a GP with zero mean and a squared exponential kernel $
    \text{Cov}(\mathcal{X}_n(v_1), \mathcal{X}_n(v_2)) = 0.01  \exp\left(-15  ||\mathcal{X}_n(v_1) - \mathcal{X}_n(v_2) ||_2^2\right)$.
\noindent
\subsubsection{Coefficient tensor generation:} We considered two approaches for generating the coefficient tensor $\Theta$: i) $\Theta$ was generated using a CP decomposition, yielding a coefficient tensor that admits a low-rank representation as in Model~(\ref{eq:tensor}), with the rank and factor matrices treated as unknown. ii) $\Theta$ was generated using a predefined spatial configuration that does not rely on a tensor decomposition. Specifically, one triangular pyramid pattern for the small-sized and two triangular pyramid pattern for the large-sized images was embedded in the image domain, with voxels inside the shape assigned a constant value of 3 and voxels outside the shape set to 0. This construction produces a coefficient tensor with localized spatial structure and discontinuities, providing a contrasting setting to the CP-based generation.

\subsection{Competing Methods and performance metric} We compared our proposed BTOT-VC model against several benchmark methods in order to evaluate prediction and parameter estimation performance. Competing methods include: (i) tensor-on-tensor (TOT) regression model \citep{TOT}, which provides a tensor-on-tensor modeling framework but without necessarily preserving the structural information in the images; (ii) robust tensor-on-tensor model (RTOT) introduced by Lee \citep{RTOT}, with the ability to address outliers in the data; (iii) another approach (RPCA) that combines the TOT framework with robust principal component analysis \citep{RTOT}. \citep{sblf}, which links predictors and outcomes through shared latent factors while modeling spatial dependencies among voxels using GP priors; (v) a two-stage DL-IIR approach \citep{IIR} that uses basis expansions to represent images and subsequently models basis coefficients by integrating a GP within a deep kernel learning framework. Of these competing methods, TOT and SBLF are Bayesian approaches, while other methods report point estimates. Further, only the DL-IIR incorporate non-linear dependencies. None of the competing approaches incorporate patch-to-voxel mapping to predict output images. The proposed BTOT-VC approach used a patch size of $3\times 3\times 3$ for the images.

To compare these methods, we used the following performance metrics to evaluate out-of-sample prediction: Pearson correlation, and relative prediction error (RPE)  defined as $\text{RPE} := \frac{\left\| \mathcal{Y}_{\text{true}} - \mathcal{Y}_{\text{predict}} \right\|_F}{\left\| \mathcal{Y}_{\text{true}} \right\|_F},$,
where $\mathcal{Y}_{\text{true}}$ is the ground truth outcome tensor and $\mathcal{Y}_{\text{predict}}$ is the predicted tensor. For the proposed method, coefficient estimation for the identifiable coefficients  ($\Theta \odot \mathcal{M}_{n,\cdot}(\mathcal{X}_{\mathcal{P},n})$) was also assessed by computing the RPE for each voxel and averaging across all voxels. However, it was not possible to report the parameter estimation under competing methods since the software used to implement them did not readily output parameter estimates. Further, we reported computation times, and additionally assessed MCMC convergence using Geweke’s diagnostic for Bayesian methods~\citep{geweke1992evaluating}.

\subsection{Results}
Results for 3D simulations are presented in Table \ref{tab:combined_3D}. Unfortunately, the RTOT, RPCA and TOT models failed to run for the 3D images of size $32 \times 32 \times 32$ due to scalability and computational constraints. As a result, we report performance metrics only for the BTOT-VC and DL-IIR models for the 3-D simulations. To facilitate a broader comparison with various competing approaches that are not scalable to large-sized 3D images, we also presented results on smaller 3D images ($8 \times 8 \times 8$) in Table~\ref{tab:rpe_3d_8cube}. 
\begin{table}[H]
\centering
\caption{ Comparison of coefficient estimation and prediction (using RPE in the format of mean(SD)) across different methods based on large-sized 3-D images ($32 \times 32 \times 32$) under different simulation settings. Results are averaged over replicates and variability is reported as standard deviation in parenthesis. Only results for the proposed approach and DL-IIR method are presented since the other competing approaches were not scalable. The better-performing model is highlighted in bold. 
}
\label{tab:combined_3D}
\scriptsize
\begin{tabular}{lllll}
\toprule
\multirow{2}{*}{\textbf{Scenario}} & \multicolumn{2}{c}{\textbf{RPE}} & \multicolumn{2}{c}{\textbf{Correlation}} \\
\cmidrule(r){2-3}  \cmidrule(r){4-5} 
&\textbf{BTOT-VC} & \textbf{DL-IIR} & \textbf{BTOT-VC} & \textbf{DL-IIR} \\
\midrule
1.a.i                     & \textbf{0.519 (0.001)} & 0.946 (0.008) & \textbf{0.855 (0.001)} & 0.688 (0.035) \\
1.a.ii                    & \textbf{0.519 (0.005)} & 0.947 (0.005) & \textbf{0.855 (0.003)} & 0.690 (0.027) \\
1.b.i                     & \textbf{0.552 (0.006)} & 0.780 (0.031) & \textbf{0.835 (0.004)} & 0.669 (0.021) \\
1.b.ii                    & \textbf{0.547 (0.004)} & 0.768 (0.040) & \textbf{0.837 (0.003)} & 0.682 (0.015) \\
1.c.i                     & \textbf{0.516 (0.001)} & 0.949 (0.005) & \textbf{0.857 (0.001)} & 0.691 (0.031) \\
1.c.ii                    & \textbf{0.504 (0.006)} & 0.946 (0.005) & \textbf{0.864 (0.004)} & 0.702 (0.014) \\
1.d.i                     & \textbf{0.519 (0.024)} & 0.912 (0.011) & \textbf{0.861 (0.010)} & 0.718 (0.025) \\
1.d.ii                    & \textbf{0.493 (0.011)} & 0.914 (0.008) & \textbf{0.872 (0.006)} & 0.712 (0.024) \\
\hline
2.a.i                     & \textbf{0.511 (0.001)} & 0.946 (0.007) & \textbf{0.860 (0.001)} & 0.702 (0.028) \\
2.a.ii                    & \textbf{0.495 (0.006)} & 0.944 (0.010) & \textbf{0.869 (0.004)} & 0.745 (0.038) \\
2.b.i                     & \textbf{0.546 (0.001)} & 0.973 (0.535) & \textbf{0.837 (0.001)} & 0.675 (0.022) \\
2.b.ii                    & \textbf{0.546 (0.005)} & 0.760 (0.036) & \textbf{0.838 (0.003)} & 0.704 (0.023) \\
2.c.i                     & \textbf{0.514 (0.008)} & 0.946 (0.007) & \textbf{0.858 (0.004)} & 0.712 (0.017) \\
2.c.ii                    & \textbf{0.507 (0.004)} & 0.947 (0.005) & \textbf{0.862 (0.003)} & 0.728 (0.024) \\
2.d.i                     & \textbf{0.600 (0.029)} & 0.903 (0.010) & \textbf{0.815 (0.012)} & 0.735 (0.024) \\
2.d.ii                    & \textbf{0.551 (0.012)} & 0.912 (0.008) & \textbf{0.839 (0.007)} & 0.725 (0.019) \\
\hline
3.a.i                     & \textbf{0.631 (0.002)} & 0.944 (0.008) & \textbf{0.776 (0.002)} & 0.698 (0.023) \\
3.a.ii                    & \textbf{0.637 (0.004)} & 0.938 (0.006) & \textbf{0.769 (0.004)} & 0.691 (0.016) \\
3.b.i                     & \textbf{0.551 (0.002)} & 0.761 (0.064) & \textbf{0.835 (0.001)} & 0.728 (0.019) \\
3.b.ii                    & \textbf{0.552 (0.003)} & 0.788 (0.074) & \textbf{0.833 (0.002)} & 0.702 (0.019) \\
3.c.i                     & \textbf{0.632 (0.002)} & 0.943 (0.007) & \textbf{0.775 (0.002)} & 0.703 (0.020) \\
3.c.ii                    & \textbf{0.634 (0.004)} & 0.941 (0.007) & \textbf{0.771 (0.003)} & 0.682 (0.019) \\
3.d.i                     & \textbf{0.695 (0.009)} & 0.913 (0.012) & \textbf{0.733 (0.003)} & 0.683 (0.027) \\
3.d.ii                    & \textbf{0.704 (0.017)} & 0.899 (0.011) & \textbf{0.726 (0.009)} & 0.671 (0.014) \\
\hline
4.a.i                     & \textbf{0.461 (0.001)} & 0.961 (0.006) & \textbf{0.891 (0.001)} & 0.580 (0.019) \\
4.a.ii                    & \textbf{0.466 (0.002)} & 0.961 (0.007) & \textbf{0.890 (0.001)} & 0.571 (0.023) \\
4.b.i                     & \textbf{0.414 (0.001)} & 0.734 (0.134) & \textbf{0.911 (0.001)} & 0.777 (0.027) \\
4.b.ii                    & \textbf{0.414 (0.002)} & 0.853 (0.128) & \textbf{0.910 (0.001)} & 0.746 (0.022) \\
4.c.i                     & \textbf{0.458 (0.001)} & 0.957 (0.007) & \textbf{0.892 (0.000)} & 0.585 (0.019) \\
4.c.ii                    & \textbf{0.463 (0.002)} & 0.957 (0.007) & \textbf{0.892 (0.001)} & 0.585 (0.014) \\
4.d.i                     & \textbf{0.487 (0.013)} & 0.931 (0.007) & \textbf{0.875 (0.006)} & 0.579 (0.019) \\
4.d.ii                    & \textbf{0.498 (0.018)} & 0.922 (0.009) & \textbf{0.871 (0.008)} & 0.516 (0.021)\\
\bottomrule
\end{tabular}
\end{table}


\begin{table}[H]
\captionsetup{skip=1pt}
\centering
\caption{Comparison of parameter estimation and out-of-sample prediction (computed using RPE in the format of mean(SD)) across different models for small sized 3D images ($8\times 8\times 8$) in various simulation scenarios. The best prediction performance is obtained under the proposed approach for all settings. }
\scriptsize
\label{tab:rpe_3d_8cube}
\begin{tabular}{cccccc}
\toprule
\textbf{Scenario} & \textbf{BTOT-VC} & \textbf{DL-IIR} & \textbf{RTOT} & \textbf{RPCA} & \textbf{TOT} \\
\midrule
1.a.i    & \textbf{0.499 (0.006)} & 0.891 (0.015) & 1.001 (0.003) & 1.001 (0.004) & 1.005 (0.007) \\
1.a.ii   & \textbf{0.55 (0.011)}  & 0.912 (0.012) & 0.999 (0.001) & 0.999 (0.003) & 1.004 (0.005) \\
1.b.i    & \textbf{0.558 (0.007)} & 0.781 (0.058) & 0.722 (0.023) & 0.934 (0.040)  & 0.978 (0.013) \\
1.b.ii   & \textbf{0.612 (0.006)} & 0.823 (0.020) & 0.888 (0.034) & 0.973 (0.018) & 0.984 (0.012) \\
1.c.i    & \textbf{0.493 (0.006)} & 0.897 (0.021) & 1.001 (0.002) & 1.002 (0.004) & 1.002 (0.002) \\
1.c.ii   & \textbf{0.532 (0.011)} & 0.916 (0.018) & 1.000 (0.001) & 0.999 (0.002) & 1.004 (0.006) \\
1.d.i    & \textbf{0.592 (0.121)} & 2.027 (1.257) & 1.020(0.001)  & 1.001 (0.008) & 1.001 (0.003) \\
1.d.ii   & \textbf{0.409 (0.026)} & 1.241 (0.570) & 0.999 (0.001) & 1.003 (0.004) & 1.004 (0.006) \\
\hline
2.a.i    & \textbf{0.414 (0.004)} & 0.878 (0.022) & 1.001 (0.003) & 1.003 (0.004) & 1.001 (0.002) \\
2.a.ii   & \textbf{0.351 (0.007)} & 0.862 (0.019) & 1.003 (0.001) & 1.001 (0.002) & 1.003 (0.002) \\
2.b.i    & \textbf{0.417 (0.005)} & 0.655 (0.093) & 0.724 (0.049) & 0.948 (0.029) & 0.953 (0.025) \\
2.b.ii   & \textbf{0.412 (0.003)} & 0.722 (0.074) & 0.793 (0.064) & 0.974 (0.016) & 0.972 (0.027) \\
2.c.i    & \textbf{0.424 (0.009)} & 0.865 (0.021) & 1.001 (0.007) & 1.001 (0.004) & 1.004 (0.004) \\
2.c.ii   & \textbf{0.365 (0.006)} & 0.871 (0.016) & 1.001 (0.002) & 1.001 (0.004) & 1.002 (0.004) \\
2.d.i    & \textbf{0.812 (0.236)} & 1.715 (1.274) & 1.100(0.001)  & 1.003 (0.008) & 1.003 (0.003) \\
2.d.ii   & \textbf{0.394 (0.054)} & 2.298 (1.270) & 1.001 (0.002) & 1.002 (0.004) & 1.002 (0.003) \\
\hline
3.a.i    & \textbf{0.634 (0.007)} & 0.889 (0.016) & 1.002 (0.005) & 1.001 (0.004) & 1.004 (0.006) \\
3.a.ii   & \textbf{0.663 (0.006)} & 0.900 (0.030) & 0.963 (0.002) & 0.965 (0.005) & 0.965 (0.005) \\
3.b.i    & \textbf{0.504 (0.004)} & 0.703 (0.085) & 0.590 (0.144) & 0.910 (0.069) & 0.916 (0.046) \\
3.b.ii   & \textbf{0.490 (0.004)} & 0.738 (0.068) & 0.545 (0.008) & 0.906 (0.043) & 0.906 (0.061) \\
3.c.i    & \textbf{0.631 (0.005)} & 0.890 (0.012) & 0.998 (0.002) & 1.003 (0.008) & 0.999 (0.008) \\
3.c.ii   & \textbf{0.668 (0.009)} & 0.893 (0.019) & 0.964 (0.003) & 0.966 (0.006) & 0.966 (0.007) \\
3.d.i    & \textbf{0.687 (0.057)} & 1.994 (1.235) & 1.002 (0.002) & 1.001 (0.005) & 1.001 (0.008) \\
3.d.ii   & \textbf{0.564 (0.031)} & 1.987 (1.830) & 0.968 (0.005) & 0.973 (0.001) & 0.965 (0.013) \\
\hline
4.a.i    & \textbf{0.464 (0.006)} & 0.923 (0.007) & 0.999 (0.004) & 1.001 (0.003) & 1.002 (0.002) \\
4.a.ii   & \textbf{0.425 (0.007)} & 0.962 (0.003) & 1.001 (0.001) & 1.002 (0.003) & 1.003 (0.001) \\
4.b.i    & \textbf{0.411 (0.003)} & 0.681 (0.055) & 0.574 (0.092) & 0.910 (0.061) & 0.955 (0.073) \\
4.b.ii   & \textbf{0.334 (0.006)} & 0.641 (0.109) & 0.523 (0.006) & 0.863 (0.025) & 0.856 (0.083) \\
4.c.i    & \textbf{0.461 (0.006)} & 0.926 (0.010) & 0.998 (0.001) & 1.001 (0.004) & 1.002 (0.002) \\
4.c.ii   & \textbf{0.422 (0.007)} & 0.968 (0.005) & 0.999 (0.001) & 1.001 (0.001) & 1.005 (0.003) \\
4.d.i    & \textbf{0.532 (0.033)} & 2.242 (1.417) & 1.001 (0.002) & 1.001 (0.005) & 1.001 (0.005) \\
4.d.ii   & \textbf{0.377 (0.034)} & 2.160 (1.324) & 0.998 (0.001) & 0.996 (0.003) & 1.003 (0.007)\\
\bottomrule
\end{tabular}%

\end{table}

From the results in Table ~\ref{tab:combined_3D}, it is evident that the BTOT-VC approach has significantly improved prediction compared to the DL-IIR approach for all scenarios for the $32\times 32 \times 32$ images. In fact, the RPE under the proposed approach is often 20-50\% lower than the DL-IIR method, resulting in orders of magnitude reduction in prediction errors.  

Other competing approaches were not computationally scalable to large-sized 3-D images.  The TOT and RTOT methods become prohibitively expensive for large-sized 3-D images due to the large number of parameters included in the model by construction. Notably, previous applications of RTOT have been typically limited to lower-dimensional settings \citep{RTOT}, reflecting the difficulty of scaling this method to large-sized 3D neuroimaging data. Moreover, the DL-IIR approach relies on a SVGD algorithm for approximate posterior inference, which is computationally fast but is limited in its ability to capture uncertainty.

For the small 3-D images ($8\times 8\times 8$), we again found that the proposed approach has a significantly lower prediction RPE compared to all other methods for all scenarios. The coefficient estimation under the proposed approach is also fairly accurate. 

The other competing approaches have less viable performance, often exceeding or nearing RPE = 1 that indicates poor prediction. The RTOT and TOT approaches specify a relatively large number of tensor coefficients by construction, which may result in overfitting and impact predictive performance. The RPCA method uses principal components derived from the images for prediction, which potentially results in information loss and results in poor predictive performance. For small-sized images, the DL-IIR approach registered considerably poor performance with the RPE often exceeding one. Further, the DL-IIR performance seems somewhat unstable with the RPE varying prominently across replicates for certain simulation scenarios (results not included due to space constraints).  

The average computing time per 1000 MCMC under the BTOT-VC model is 87.594 mins for large-sizes images and 5.670 min for small-sized images, using 4 CPU cores and 100 GB memory. The corresponding computation times under competing approaches for small-sized images were 5.670 min for BTOT-VC, and 15.868  min for TOT, per 1,000 iterations. Computation time can be further reduced by lowering the number of iterations, as MCMC convergence is typically rapid. 
In all cases, the absolute Geweke z-scores corresponding to the identifiable coefficient terms $\Theta(v) \mathcal{M}_{n,v}(\mathcal{X}_{n, \text{patch}}(v))$ were below 1.96, indicating satisfactory mixing and convergence. This is verified via MCMC traceplots presented for a few voxels in \ref{s:supp} Figure \ref{fig:traceplot}. Overall, the proposed approach consistently delivers strong predictive performance, results in model parsimony and robust parameter estimation, and is scalable to large-dimensional 3-D imaging data. For each image, tensor ranks $R$ from 1 to 10 were considered. For simplicity, the ranks of all coefficient tensors were constrained to be equal. The optimal rank was selected by minimizing the DIC, as described in Section~\ref{s:addi} Figure~\ref{fig:dic}. And the MCMC diagnosis are in \ref{s:addi} with traceplot in Figure~\ref{fig:traceplot} and the Geweke $z$-score summary in Table \ref{tab:geweke}.

\section{ADNI Application}
\label{sec:ADNI}
\subsection{Data Description}
To evaluate the practical utility of our proposed framework, we applied it to longitudinal MRI data from the Alzheimer’s Disease Neuroimaging Initiative (ADNI) (adni.loni.usc.edu). The dataset includes 639 participants with T1-weighted structural MRI scans available at three timepoints: baseline (bl), follow-up at 6 month, and follow up at 12 month visits. This cohort comprises 195 cognitively normal healthy control (HC) subjects, 311 participants diagnosed with MCI, and 133 individuals with AD at baseline. Apart from imaging data, four baseline covariates were considered, including demographic and clinical factors: age, gender (encoded as 0 for female and 1 for male), APOE4 genotype (number of $\varepsilon4$ alleles: 0, 1, or 2), and years of education. We performed the analysis separately for the AD group, as well as the MCI group. For the competing methods, we first fit a voxel-wise regression approach to remove the effects of the baseline covariates from the outcome images, and subsequently used these residuals for fitting the tensor-on-tensor regression models. The data for cognitive normal controls was used for downstream brain age calculations (not directly used for image-on-image regression) as described in the sequel. We provide a table to describe the demographic details of the dataset in Table \ref{tab:demo}.
\noindent
\begin{table}[p]
\centering
\scriptsize
\caption{Demographic and clinical characteristics of AD and MCI Groups}
\label{tab:demo}
\begin{tabular}{lcc}
\hline
 & \textbf{AD group} & \textbf{MCI group} \\
\hline
\textbf{N} & 133 & 311 \\[0.3em]

\textbf{Age (Mean (SD))} & 75.8 (7.59) & 75.9 (7.06) \\

\textbf{Years of Education (Mean (SD))} & 14.7 (3.11) & 15.7 (3.00) \\[0.6em]

\textbf{Gender} &  &  \\
\quad Female & 64 (48.1\%) & 110 (35.4\%) \\
\quad Male   & 69 (51.9\%) & 201 (64.6\%) \\[0.6em]

\textbf{APOE4} &  &  \\
\quad 0 & 44 (33.1\%) & 143 (46.0\%) \\
\quad 1 & 58 (43.6\%) & 131 (42.1\%) \\
\quad 2 & 31 (23.3\%) & 37 (11.9\%) \\
\hline
\end{tabular}
\label{tab:demo}
\end{table}

\subsubsection{T1w-MRI pre-processing:
} All T1-weighted images were preprocessed using the Advanced Normalization Tools (ANTs) longitudinal registration pipeline \citep{tustison2010n4itk}. This pipeline spatially normalized each scan to a population-based template, which was constructed from 52 cognitively normal subjects from the ADNI-1 cohort and distributed by the ANTs group \citep{tustison2019longitudinal}. The preprocessing steps included: (i) N4 bias field correction \citep{tustison2014large}, which mitigates intensity non-uniformity and improves tissue contrast; (ii) Symmetric diffeomorphic registration \citep{avants2008symmetric}, used to align each subject’s image to the template space with topological consistency; and (iii) 6-tissue segmentation that yielded subject-specific masks for CSF, gray matter (GM), white matter (WM), deep gray matter (DGM), brainstem, and cerebellum. CT maps were then estimated using the Diffeomorphic Registration-based Cortical Thickness method. For computational efficiency and to enable voxel-wise prediction across longitudinal timepoints, all 3D images were isotropically downsampled to a uniform spatial dimension of $48 \times 48 \times 48$. The ADNI images were collected across different sites/scanners where harmonized using Combat \citep{beer2020longitudinal}.

\noindent

\subsection{Analysis Outline}

Our analysis has two main objectives. First, we aim to predict CT images at month 12 using baseline and month 6 data and evaluate prediction accuracy. Specifically, we train a tensor-on-tensor regression model with baseline and month 6 images as inputs and outputs, then predict month 12 images using month 6 data. This framework provides external validation of the model’s ability to forecast future brain changes from earlier imaging, enabling prediction of neurodegeneration and identification of early neurobiological mechanisms linked to future AD progression. This analysis is biologically motivated, as AD is believed to develop decades before clinical diagnosis \citep{dubois2016preclinical}. Second, we conduct a downstream analysis using the predicted month 12 images to estimate the predicted brain age gap (BAG) and compare it with the actual BAG derived from observed month 12 scans. The BAG, defined as the difference between predicted brain age and chronological age, serves as a marker of structural brain aging; a positive BAG indicates accelerated aging linked to greater AD severity and earlier symptom onset, even at the MCI stage \citep{driscoll2009longitudinal}, \citep{wang2019gray}. Prior studies have shown that BAG reflects regional atrophy and correlates with clinical progression \citep{franke2012longitudinal}, \citep{lowe2016effect}, supporting its value as a biomarker of neurodegeneration. Accurate BAG forecasting from early longitudinal data could improve early AD detection and optimize patient recruitment in phase II and III trials, where late enrollment has contributed to high failure rates \citep{kim2022alzheimer}. To account for disease-stage–specific progression patterns, the analysis was conducted separately for AD and MCI groups, hereafter referred to as the AD Long and MCI Long analyses, respectively.\\

To further assess the generalizability and predictive robustness of the proposed BTOT-VC framework, we conducted an additional out-of-sample prediction analysis based solely on month 6 imaging data. In this analysis, subjects with month 6 and month 12 scans were randomly split into training and test sets with a ratio of 75:25. The model was trained using month 6 images from the training set as inputs and the corresponding observed month 12 images as responses, and subsequently evaluated by predicting month 12 images for held-out test subjects using their month 6 scans only. Prediction accuracy was quantified using voxel-wise and ROI-level metrics, providing an out-of-sample evaluation of the model’s ability to forecast neuro-anatomical changes. This cross-sectional prediction setting complements the longitudinal forecasting analysis described above by explicitly separating training and testing cohorts, thereby offering a stringent assessment of external predictive performance. The predicted month 12 images obtained from this external analysis were then used in the same downstream BAG analysis as described previously, enabling a direct comparison between BAG estimated from predicted images and BAG derived from observed month 12 scans. This out-of-sample prediction analysis was also conducted separately for AD and MCI cohorts, referred to as the AD Out-of-Sample and MCI Out-of-Sample settings, respectively. Together, these analyses provide complementary evidence of the BTOT-VC model’s capacity to capture biologically meaningful patterns of brain aging and to generalize across subjects in both longitudinal and cross-sectional prediction settings.\\

To facilitate comparison with competing approaches that are not scalable to high-dimensional images, we performed the tensor-on-tensor regression modeling separately for each region of interest (ROI) as defined under the 83-region Desikan–Killiany–Tourville (DKT) cortical parcellation atlas \citep{DKT}. This atlas, originally  defined in the MNI152NLin6 standard anatomical space at a $1 mm^3$ isotropic resolution, was nonlinearly registered to the study-specific population template using the same ANTs deformation fields to ensure anatomical consistency across subjects. Following registration, both the CT images and the atlas were spatially cropped to the cortical region of interest and downsampled to a uniform $48 \times 48 \times 48$ voxel grid. For the BAG analysis, the model for computing BAG was trained using ROI level data from the DKT atlas as inputs and chronological age as the scalar outcome, using data from cognitively normal controls, as per convention. The BAG is calculated as the difference between the predicted brain age and the chronological age. We use both random forest (RF) model to train the BAG prediction model. Subsequently, this trained model was used to predict BAG for AD and MCI individuals based on their predicted CT images at month 12, and the predicted BAG was compared with the actual BAG computed based on observed images at month 12. To mitigate regression-to-the-mean effects, we applied an age-bias correction and used the bias-adjusted BAG values \citep{BAG_correction}.

To further demonstrate the advantage of the proposed BTOT-VC framework in modeling brain aging trajectories, we conducted an additional BAG prediction comparative analysis using images at early visit directly to predict the BAG in the future. Specifically, we compared BAG estimates derived from ROI features extracted from BTRR-predicted images with BAG predictions obtained directly from baseline images without intermediate image forecasting. In the out-of-sample setting, the full sample was randomly split into training and test sets using a 75:25 ratio and repeated across 10 replicates, with identical splits applied to both approaches to ensure fair comparison. In both cases, a random forest model was used to predict BAG from ROI-level features. To establish a reference (“ground truth”) BAG, we trained the brain age prediction model using HC subjects and then applied it to observed month 12 images from the AD and MCI cohorts to estimate brain age. The true BAG was defined as the difference between the estimated brain age and the subject’s chronological age. This design allows us to directly assess whether incorporating BTRR-based image prediction improves downstream BAG estimation relative to baseline-only approaches. We denote this analysis as Baseline Prediction

\subsection{Results}
\subsubsection{Cortical thickness MRI prediction}
\begin{figure}[p]
\centering
\begin{subfigure}{0.7\textwidth}
  \centering
  \includegraphics[width=\linewidth]{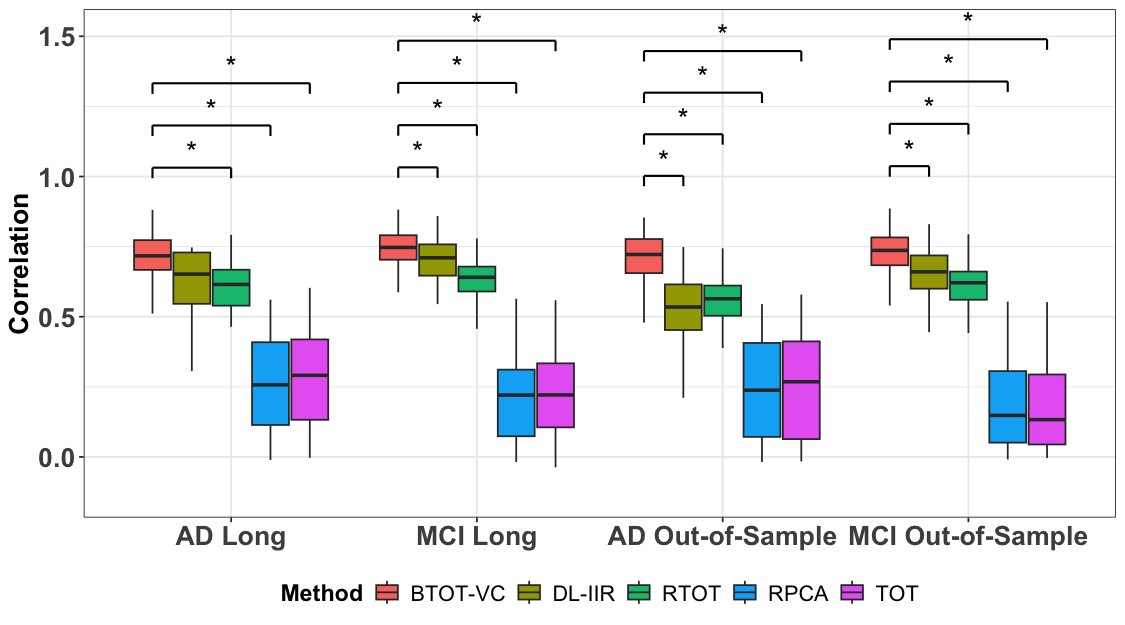}
  \caption{}
  \label{fig:sub:a}
\end{subfigure}
\hfill
\begin{subfigure}{0.7\textwidth}
  \centering
  \includegraphics[width=\linewidth]{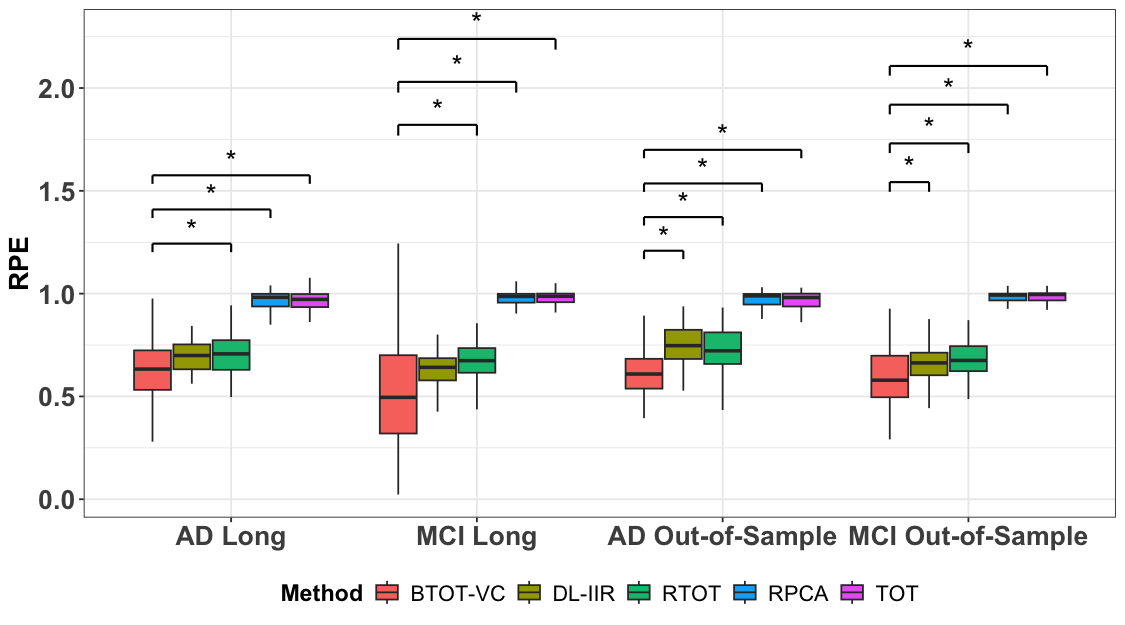}
  \caption{}
  \label{fig:sub:b}
\end{subfigure}

\vspace{0.5em}

\caption{\textbf{ ROI-Level Longitudinal Prediction Accuracy.} 
    Panels (a)– (b) summarize voxel-wise prediction accuracy for cortical thickness across 83 ROIs, using correlation (a), RPE (b). For each ROI, boxplots display the distribution of correlation and RPE values across voxels. Cortical thickness values were extracted using the DKT atlas, and all models were trained and evaluated independently within each group and ROI.}
    \label{fig:ADNI_res}
\end{figure}

Prediction accuracy was evaluated voxel-wise within each ROI using Pearson correlation, and RPE, and the results are summarized in Figure \ref{fig:ADNI_res} (a) - (b) and Table \ref{tab:ADNI_ROI}. These plots demonstrate that our proposed BTOT-VC model consistently achieves superior voxel-wise prediction performance across all clinical groups and ROIs. Specifically, our model yields the lowest RPE, and the highest correlation values in the vast majority of ROIs within each sub-group considered for analysis. Improvements in RPE were statistically significant relative to all competing approaches across cohorts, with the exception of the DL-IIR model in the AD longitudinal setting, where performance was comparable. However, this apparent similarity should be interpreted in light of model coverage: while BTOT-VC produced valid predictions for all 83 cortical ROIs, the DL-IIR model yielded predictions for only 13 ROIs. Consequently, although DL-IIR achieves competitive accuracy on this limited subset, its substantially reduced spatial coverage restricts its practical utility. Therefore, our results indicate the strong ability of the proposed approach to accurately forecast future brain structural patterns based on earlier MRI scans. These findings underscore the effectiveness of our Bayesian tensor-on-tensor approach in capturing spatial dependencies in high-dimensional neuroimaging data across longitudinal visits, enabling interpretable and biologically meaningful prediction of future neurobiological changes at a fine-grained voxel level.\\
\noindent

\begin{table}[p]
\centering
\scriptsize
\caption{Summary of predictive performance across methods and study groups. For each method, we report the voxel-wise prediction Pearson correlation and RPE, summarized across ROIs in the format of mean(SD), under three evaluation settings: AD longitudinal, MCI longitudinal, and external AD validation.}
\label{tab:ADNI_ROI}
\begin{tabular}{llccc}
\hline
\textbf{Group} & \textbf{Method} & \textbf{Correlation} & \textbf{RPE} & \makecell{\textbf{Count of} \\ \textbf{Predictable ROIs}} \\
\hline\\
\multirow{5}{*}{\textbf{AD Long}}
 & \textbf{BTOT-VC} & \textbf{0.715 (0.090)} & \textbf{0.634 (0.151)} & 83 \\
 & \textbf{DL-IIR}  & 0.587 (0.157) & 0.712 (0.114) & 13 \\
 & \textbf{RTOT}    & 0.609 (0.083) & 0.703 (0.100) & 83 \\
 & \textbf{RPCA}    & 0.256 (0.165) & 0.966 (0.041) & 83 \\
 & \textbf{TOT}     & 0.276 (0.174) & 0.960 (0.049) & 83 \\
[0.6em]
\multirow{5}{*}{\textbf{MCI Long}}
 & \textbf{BTOT-VC} & \textbf{0.731 (0.091)} & \textbf{0.584 (0.351)} & 83 \\
 & \textbf{DL-IIR}  & 0.682 (0.131) & 0.667 (0.276) & 79 \\
 & \textbf{RTOT}    & 0.635 (0.083) & 0.673 (0.088) & 82 \\
 & \textbf{RPCA}    & 0.211 (0.154) & 0.975 (0.040) & 82 \\
 & \textbf{TOT}     & 0.219 (0.157) & 0.974 (0.044) & 82 \\
[0.6em]
\multirow{5}{*}{\textbf{AD Out-of-Sample}}
 & \textbf{BTOT-VC} & \textbf{0.697 (0.118)} & \textbf{0.619 (0.146)} & 83 \\
 & \textbf{DL-IIR}  & 0.536 (0.113) & 0.753 (0.105) & 82 \\
 & \textbf{RTOT}    & 0.557 (0.095) & 0.731 (0.099) & 83 \\
 & \textbf{RPCA}    & 0.241 (0.172) & 0.968 (0.039) & 83 \\
 & \textbf{TOT}     & 0.260 (0.183) & 0.963 (0.046) & 83 \\
\multirow{5}{*}{\textbf{MCI Out-of-Sample}}
 & \textbf{BTOT-VC} & \textbf{0.724 (0.098)} & \textbf{0.619 (0.146)} & 83 \\
 & \textbf{DL-IIR}  & 0.650 (0.099) & 0.680 (0.181) & 79 \\
 & \textbf{RTOT}    & 0.618 (0.086) & 0.679 (0.090)) & 83 \\
 & \textbf{RPCA}    & 0.194 (0.160) & 0.979 (0.035) & 83 \\
 & \textbf{TOT}     & 0.186 (0.158) & 0.980 (0.039) & 83 \\

\hline
\end{tabular}
\end{table}
\noindent

\subsubsection{Part 2: Brain age prediction using predicted scans} 
To assess the accuracy of BAG estimation, we evaluate the agreement between the predicted BAG and the actual BAG values in terms of Pearson correlation and root mean squared error (RMSE). We also evaluated the ability of the proposed tensor-on-tensor approaches to predict accelerated brain aging (BA) that is defined using the sign of the BAG (accelerated BA if BAG$>0$ vs.\ non-accelerated BA if BAG$\le 0$). We evaluate the concordance between the predicted and actual BAG signs using the F1 score, with a higher F1 score indicating increased ability to predict accelerated BA.  The F1 score is defined as the harmonic mean of precision and recall, $\text{F1} = \frac{2(\text{precision}\times \text{recall})}{\text{precision} + \text{recall}}$, and provides a balanced measure of classification performance, particularly when class distributions are imbalanced. We present the results corresponding to a random forest model that was used to train the brain age prediction model using ROI-level neuroimaging features as inputs (the results for the XGBoost based training model is quite similar and not presented here).
Our results reveal elevated BAG values in AD and MCI groups compared to controls, that is expected due to potentially accelerated brain aging resulting from AD progression. The RMSE and F1 score results for different approaches are shown in Table \ref{tab:bag_pref}. For all the MCI subcohorts, the proposed BTOT-VC method had improved BAG estimation and considerably greater F1 score indicating its ability to accurately predict accelerated brain aging at future time points. For the AD group, the BAG estimation performance of the proposed approach was comparable to other approaches. Comparisons with Baseline Prediction emphasize the importance of incorporating MRI image forecasting in modeling brain aging trajectories. BAG prediction based solely on early-visit MRI yielded inferior performance relative to methods leveraging predicted month-12 images, with the largest degradation observed in the AD cohort. This result underscores the importance of explicitly modeling future structural brain changes when estimating brain aging trajectories in progressive neurodegenerative disease. Overall, our findings suggest that voxel-wise image prediction with spatially informed modeling and incorporating low-rank structures for the tensor coefficients can be successfully used to predict future spatially distributed neurobiological changes and identify individuals who are most likely to experience accelerated brain aging in the future. 

\noindent

\begin{table}[p]
\centering
\caption{\textbf{Prediction and classification performance of brain age gap (BAG) across study groups and modeling strategies.} Results are reported separately for AD and MCI cohorts under three evaluation settings. (i) Longitudinal forecasting (AD long, MCI long): month-12 MRI images are predicted using month-6 images using the whole cohort, and BAG is computed from the predicted images. (ii) Out-of-sample forecasting (AD Out-of-sample, MCI-Out-of-sample): month-12 images are predicted from month-6 images using independent training and test splits (75:25), providing external validation of image forecasting performance. (iii) Early-visit prediction (Early-Visit Prediction): BAG is predicted directly from early-visit MRI without image forecasting, using either baseline or month-6 images.For each method, we report BAG prediction correlation and root mean squared error (RMSE), along with classification performance for accelerated brain aging (BAG $> 0$).
}
\label{tab:bag_pref}
\scriptsize
\begin{tabular}{llccc}
\hline
\textbf{Group} & \textbf{Method} & \textbf{Correlation} & \textbf{RMSE} & \textbf{F1 Score} \\
\hline
\multirow{5}{*}{\textbf{AD Long}}
 & \textbf{BTOT-VC} & \textbf{0.816} & \textbf{2.398} & \textbf{0.923} \\
 & \textbf{DL-IIR}  & 0.764 & 3.398 & 0.873 \\
 & \textbf{RTOT}    & 0.809 & 2.901 & 0.788 \\
 & \textbf{RPCA}    & 0.811 & 2.515 & 0.842 \\
 & \textbf{TOT}     & 0.806 & 2.505 & 0.842 \\
[0.6em]

\multirow{5}{*}{\textbf{MCI Long}}
 & \textbf{BTOT-VC} & \textbf{0.817} & \textbf{1.764} & \textbf{0.889} \\
 & \textbf{DL-IIR}  & 0.751 & 2.369 & 0.792 \\
 & \textbf{RTOT}    & 0.704 & 2.425 & 0.801 \\
 & \textbf{RPCA}    & 0.691 & 2.389 & 0.845 \\
 & \textbf{TOT}     & 0.705 & 2.335 & 0.817 \\
[0.6em]

\multirow{5}{*}{\textbf{AD Out-of-Sample}}
 & \textbf{BTOT-VC} & \textbf{0.980 (0.004)} & \textbf{1.844 (0.379)} & \textbf{0.953 (0.027)} \\
 & \textbf{DL-IIR}  & 0.963 (0.008) & 3.180 (0.185) & 0.856 (0.019) \\
 & \textbf{RPCA}    & 0.970 (0.007) & 1.994 (0.288) & 0.933 (0.025) \\
 & \textbf{RTOT}    & 0.964 (0.006) & 2.792 (0.225) & 0.893 (0.047) \\
 & \textbf{TOT}     & 0.970 (0.006) & 1.886 (0.207) & 0.949 (0.020) \\

\multirow{5}{*}{\textbf{MCI Out-of-Sample}}
 & \textbf{BTOT-VC} & 0.966 (0.006) & 2.464 (0.119) & \textbf{ 0.924 (0.020)} \\
 & \textbf{DL-IIR}  & \textbf{0.969 (0.004)} & 2.662 (0.170) &  0.919 (0.023)\\
 & \textbf{RPCA}    & 0.935 (0.006) & 2.973 (0.083) & 0.921 (0.018) \\
 & \textbf{RTOT}    & 0.964 (0.005) & 2.528 (0.167)& 0.915 (0.025)\\
 & \textbf{TOT}     & 0.967 (0.004)& \textbf{2.457 (0.126)}&  0.919 (0.018)\\
\multirow{4}{*}{\textbf{Early-Visit Prediction}}
  & \textbf{AD Baseline}  & -0.035 (0.117) & 11.680 (0.898) & 0.577 (0.080) \\
  & \textbf{AD Month 6}  & -0.040 (0.125) & 11.758 (0.728) &  0.600 (0.076)\\
  & \textbf{MCI Baseline}  & 0.927 (0.020) & 3.521 (0.304) & 0.804 (0.026)\\ 
  & \textbf{MCI Month 6}  & 0.935 (0.019) & 3.297 (0.277) &  0.805 (0.034)\\
\hline
\end{tabular}
\end{table}

\noindent

\section{Discussion}
\label{s:discuss}
We developed one of the first Bayesian semi-parametric approach for TOT modeling that can incorporate flexible local non-linear dependencies via voxel-wise Gaussian process specifications as well as spatial heterogeneity via a patch-to-voxel mapping. By combining low-rank tensor coefficient decompositions with parallelized computation of voxel-wise local effects, the proposed method is computationally scalable to large-dimensional 3D imaging data. Therefore, our approach addresses major limitations in TOT literature with important applications to 3D image-on-image regression, for which there is limited literature. Rigorous simulations help illustrate the superior numerical performance of the proposed approach over competing methods. Application to longitudinal AD imaging data highlights the method’s ability to forecast future neurodegenerative changes and accelerated brain aging, providing a powerful tool for early detection with the potential to inform the design of AD clinical trials.

\section{Data availability}
The data analyzed in this study were obtained from the publicly available Alzheimer’s Disease Neuroimaging Initiative (ADNI) database (https://adni.loni.usc.edu).

\section{Funding sources}
This work was made possible by generous support from National Institute on Aging (award number R01 AG071174).

\bibliographystyle{plainnat}
\bibliography{references-SK}  

\appendix
\section{Supplementary Materials}
\label{s:supp}
\subsection{Deviance Information Criteria for Tensor Rank Selection}
\label{s:dic}
Given a selected tensor rank $R$ and the model parameters $\Psi^k=\{\Gamma^k,\Theta^k,\mathcal{D}_s^k,\sigma_e^k \}$ corresponding to the $k$-th iteration, the deviance $D^k$ under Model (\ref{eq:tensor}) is expressed as
$$
D^k=D(\{\mathcal{Y}_n(v) \}_{n \in[1,N],v \in \mathcal{V}}|\Psi^k )=-2logL(\mathcal{Y}_n(v)|\Psi^k )
$$
$$
= -2log \prod_{n=1}^{N} \prod_{v \in \mathcal{V}} L(\mathcal{R}_{n}^k(v)|\sigma_e^k) = \sum_{n=1}^{N} \sum_{v \in \mathcal{V}} ((\frac{\mathcal{R}_{n}^k(v)}{\sigma_e^k})^2 + 2log \sigma_e^k + log(2\pi))
$$
where $\mathcal{R}_{n} = \mathcal{Y}_{n} - \Gamma^k - \Theta^k \odot \mathcal{M}_{n,\cdot}(\mathcal{X}_{\mathcal{P},n})^k - \sum_{s=1}^S \mathcal{D}_s^k z_{ns}$, $\mathcal{V}$ is the voxel space, and $L(\cdot)$ denotes the likelihood function. Denoting $\bar{D}$ as the mean value of $D^k$ across all post-burn-in MCMC iterations, $\bar{\Psi}$ as the posterior mean of model parameters across post-burn-in MCMC iterations, and $\{\mathcal{Y}_{n}\}$ as the set of all tensor-valued outcomes $\mathcal{Y}_{n}$ across all subjects n, and observed voxels $v$, the estimated effective number of parameters $p_{D,DIC}$ is given by $p_{D,DIC}= \bar{D} - D(\{Y_n \}\mid\bar{\Psi})$. Moreover, DIC is defined as $DIC = \bar{D} + pD$, where $\bar{D}$ penalizes poor model fits and $p_{D,DIC}$ penalizes models with a large number of parameters resulting from large selected rank $R$ \citep{shriner2009deviance}. The Bayesian model is fit repeatedly for different choices of the tensor rank, and the optimal rank is chosen corresponding to the model that returns the lowest DIC score.
\subsection{Posterior Computation Steps}
\label{s:posterior_computation}
Suppose we have $N_{\text{train}}$ subjects in the training set and $N_{\text{test}}$ subjects in the test set, and a total of $N$ subjects. We use $\circ$ to denote the outer product. For a given dimension $d \in [1,D]$ and tensor margin index $j \in [1,p_d]$, let $\mathcal{Y}_{n,\cdot,dj}$ be the slice of $\mathcal{Y}_{n,\cdot}$ obtained by fixing dimension $d$ at index $j$ as described in section \ref{sec:Notations}, and let a scalar product between two tensor slices $\mathcal{A}_{\cdot,dj}$ and $\mathcal{B}_{\cdot,dj}$ is defined as $\langle \mathcal{A}_{\cdot,dj}, \mathcal{B}_{\cdot,dj} \rangle = \sum_{i_1,\ldots, i_{d-1}, i_d=j, i_{d+1},\ldots i_D}\mathcal{A}_{(i_1,i_2,\ldots, i_D)} \mathcal{B}_{(i_1,i_2,\ldots, i_D)} $ that holds mode $d$ fixed at the $j$th element $(j\in \{1,\ldots,p_d \})$, while taking the product over all remaining dimensions. Denote the squared Euclidean norm of a vector as $\| \cdot \|_2^2$. Then the sampling steps for the full MCMC algorithm are as follows, where ranks $r \in [1, R]$ and dimension $d \in [1, D]$ are looped through. 

\begin{enumerate}[label=\textbf{Step \arabic*:}, leftmargin=1em, align=left]
    \item Let $\mathcal{Y}_{n,r}^{\gamma} = \mathcal{Y}_{n} - \hat{\Gamma}_r - \widehat{\Theta} \odot \widehat{\mathcal{M}_{n,\cdot}(\mathcal{X}_{\mathcal{P},n})} - \sum_{s=1}^S \widehat{\mathcal{D}_s}  z_{ns}$ be the rank $r$ specific residual corresponding to the $\Gamma$ term, where $\hat{\Gamma}_r = \sum_{r' = 1, r' \neq r}^R \hat{\boldsymbol{\gamma}}_{1\cdot,r'} \circ \cdots \circ \hat{\boldsymbol{\gamma}}_{D\cdot,r'}$ where $\hat{\boldsymbol{\gamma}}_{1\cdot,r'}, \cdots , \hat{\boldsymbol{\gamma}}_{D\cdot,r'}$ are sampled from the most recent iteration, and $\widehat{\Theta}$, $\widehat{\mathcal{D}_s}$ are taken from the most recently sampled instances of its tensor margins. And $\widehat{\mathcal{M}_{n,\cdot}(\mathcal{X}_{\mathcal{P},n})}$ is $\mathcal{M}_{n,\cdot}(\mathcal{X}_{\mathcal{P},n})$ sampled from the most recent GP based on the updated $\phi_1$ and $\phi_2$ from current iteration. The $j$-th element for tensor margin $\boldsymbol{\gamma}_{d\cdot, r}$ for $j \in \{2,\cdots, p_d-1\}$ denoted $\gamma_{d\cdot, r,j}$ follows the conditional posterior:
    $$
    \pi(\gamma_{d\cdot,r,j}|-) = \mathcal{N}(\frac{n_{d\cdot,r,j}^\gamma \tau^\gamma w_{d,r}^\gamma + e^{-\alpha_{d,r}^\gamma}(\gamma_{d\cdot,r,j-1}+\gamma_{d\cdot,r,j+1})}{m_{d\cdot,r,j}^\gamma \tau^\gamma w_{d,r}^\gamma + 1 + e^{-2\alpha_{d,r}^\gamma}}, \frac{\tau^\gamma w_{d,r}^\gamma }{m_{d\cdot,r,j}^\gamma \tau^\gamma w_{d,r}^\gamma + 1 + e^{-2\alpha_{d,r}^\gamma}})
    $$
    ,where $\bm{T}_{d\cdot,r}^\gamma = \boldsymbol{\gamma}_{1\cdot,r}\circ\cdots\circ \boldsymbol{\gamma}_{d-1\cdot,r}\circ\boldsymbol{\gamma}_{d+1\cdot,r}\circ\cdots\circ\boldsymbol{\gamma}_{D\cdot,r}$, $m_{d\cdot,r,j}^\gamma = \frac{1}{\sigma_e^2} \sum_{n=1}^{N_{\text{train}}}\sum_{v'\in \mathcal{V}_{dj}} (\bm{T}_{d\cdot,r}^\gamma(v'))^2$, 
    and $n_{d\cdot,r,j}^\gamma = \frac{1}{\sigma_e^2} \sum_{n=1}^{N_{\text{train}}} 
    \langle \mathcal{Y}_{n,r,\cdot,dj}^{\gamma}, \bm{T}_{d\cdot,r}^\gamma \rangle $. For $j \in \{1, p_d\}$, the posterior distributions are expressed as: 
\[
\pi(\gamma_{d\cdot,r,j}\mid -) =
\mathcal{N}\!\left(
\frac{n_{d\cdot,r,j}^\gamma \tau^\gamma w_{d,r}^\gamma + e^{-\alpha_{d,r}^\gamma} \gamma_{d\cdot,r,j+1}}
     {m_{d\cdot,r,j}^\gamma \tau^\gamma w_{d,r}^\gamma + 1},
\;
\frac{\tau^\gamma w_{d,r}^\gamma}
     {m_{d\cdot,r,j}^\gamma \tau^\gamma w_{d,r}^\gamma + 1}
\right),
\quad \text{for } j=1,
\]
\[
\pi(\gamma_{d\cdot,r,j}\mid -) =
\mathcal{N}\!\left(
\frac{n_{d\cdot,r,j}^\gamma \tau^\gamma w_{d,r}^\gamma + e^{-\alpha_{d,r}^\gamma} \gamma_{d\cdot,r,j-1}}
     {m_{d\cdot,r,j}^\gamma \tau^\gamma w_{d,r}^\gamma + 1},
\;
\frac{\tau^\gamma w_{d,r}^\gamma}
     {m_{d\cdot,r,j}^\gamma \tau^\gamma w_{d,r}^\gamma + 1}
\right),
\quad \text{for } j=p_d.
\]
    \item For a given dimension $d \in [1, D]$ and rank $r \in [1, R]$, the diagonal entries of the tensor margin covariance matrix $\mathbf{W}_{d,r}^\gamma$, denoted as $w_{d,r}^\gamma$, follows the generalized inverse-Gamma (gIG) posterior~\citep{mead2015generalized}

    $$
    \pi(w_{d,r}^\gamma \mid -) = \text{gIG} \left( \mu = 1 - \frac{p_d}{2}, \, \chi = \frac{c_{d,r}^\gamma}{1 - e^{-2\alpha_{d,r}^\gamma}}, \, \psi = \lambda_{d,r}^\gamma \right),
    $$
    where {\tiny$
    c_{d,r}^\gamma = \frac{1}{\tau^\gamma} \left\{ \sum_{j=2}^{p_d-1} \left( 1 + e^{-2\alpha_{d,r}^\gamma} \right) \| \gamma_{d\cdot, r, j} \|_2^2 + \| \gamma_{d\cdot, r, 1} \|_2^2 + \| \gamma_{d\cdot, r, p_d} \|_2^2 - 2e^{-\alpha_{d,r}^\gamma} \sum_{j=1}^{p_d-1} \gamma_{d\cdot, r, j}^\top \gamma_{d\cdot, r, j+1} \right\}.
    $}
    \item The rate parameter $\lambda_{d,r}^\gamma$ follows the conditional posterior {\small$\pi(\lambda_{d,r}^\gamma \mid -) = \text{Ga} \left( a_\lambda + p_d, \, b_\lambda + \frac{p_d w_{d,r}^\gamma}{2} \right).$}
    \item The global variance scale parameter $\tau^\gamma$ follows the conditional posterior 
    $$
    \pi(\tau^\gamma \mid -) = \text{gIG} \left( \mu = a_\tau - \frac{R(p_1 + \cdots + p_D)}{2}, \, \chi = \sum_{r=1}^R \sum_{d=1}^D \boldsymbol{\gamma}_{d\cdot, r}^\top (\mathbf{W}_{d,r}^\gamma)^{-1} \boldsymbol{\gamma}_{d\cdot, r}, \, \psi = 2b_\tau \right).
    $$
    \item The conditional posterior for lengthscale parameter $\alpha_{d,r}^\gamma$ satisfies the following
    $$
    \pi(\alpha_{d,r}^\gamma \mid -) \propto (\alpha_{d,r}^\gamma)^{a_\alpha - 1} (1 - e^{-2\alpha_{d,r}^\gamma})^{-\frac{1}{2}(p_d-1)} \exp \left( -\frac{1}{2} \left( \boldsymbol{\gamma}_{d\cdot, r}^\top (\mathbf{W}_{d,r}^\gamma)^{-1} \boldsymbol{\gamma}_{d\cdot, r} + 2b_\alpha \alpha_{d,r}^\gamma \right) \right).
    $$
    Because the conditional posterior for $\alpha_{d,r}^\gamma$ does not correspond to a closed-form distribution, $\alpha_{d,r}^\gamma$ is sampled using a MH step, using the proposal density $\alpha_{d,r}^{(x+1)^\gamma} \mid \alpha_{d,r}^{(x)^\gamma} \sim \text{log-Normal}(\alpha_{d,r}^{(x)^\gamma}, \sigma_\alpha^2),$ where $s_x$ indexes the MCMC iteration, and $\sigma_\alpha^2$ is fixed for all terms.

    \item For a given $s^* \in [1,S]$, let $\mathcal{Y}_{n,r,s}^{\delta} = \mathcal{Y}_{n}- \widehat{\Gamma}  - \widehat{\Theta} \odot \widehat{\mathcal{M}_{n,\cdot}(\mathcal{X}_{\mathcal{P},n})} - \sum_{s=1,s\neq s^*}^S \widehat{\mathcal{D}_{s}} z_{n{s}} - \widehat{\mathcal{D}_{s,r}} z_{n{s^*}}$ be the rank $r^*$ specific residual corresponding to the $\mathcal{D}_s$ term. Where $\widehat{\mathcal{D}_{s,r}} = \sum_{r' = 1, r' \neq r}^R \hat{\boldsymbol{\delta}}_{1\cdot,r,s'} \circ \cdots \circ \hat{\boldsymbol{\delta}}_{D\cdot,r,s'}$ where $\hat{\boldsymbol{\delta}}_{1\cdot,r,s'}, \cdots , \hat{\boldsymbol{\delta}}_{D\cdot,r,s'}$ are sampled from the most recent iteration, and $\widehat{\Gamma}$, $\widehat{\Theta}$, $\widehat{\mathcal{D}_s}$ are taken from the most recently sampled instances of its tensor margins and $\widehat{\mathcal{M}_{n,\cdot}(\mathcal{X}_{\mathcal{P},n})}$ is $\mathcal{M}_{n,\cdot}(\mathcal{X}_{\mathcal{P},n})$ sampled from the most recent GP based on the updated $\phi_1$ and $\phi_2$ from current iteration. The $j$-th element for tensor margin $\boldsymbol{\delta}_{d\cdot, r,s}$ for $j \in \{2,\cdots, p_d-1\}$ denoted $\delta_{d\cdot, r,s,j}$ follows the conditional posterior:
    $$
    \pi(\delta_{d\cdot,r,s,j}|-) = \mathcal{N}(\frac{n_{d\cdot,r,s,j}^\delta \tau_s^\delta w_{d,r,s}^\delta + e^{-\alpha_{d,r,s}^\delta}(\delta_{d\cdot,r,s,j-1}+\delta_{d\cdot,r,s,j+1})}{m_{d\cdot,r,s,j}^\delta \tau_s^\delta w_{d,r,s}^\delta + 1 + e^{-2\alpha_{d,r,s}^\delta}}, \frac{\tau_s^\delta w_{d,r,s}^\delta }{m_{d\cdot,r,s,j}^\delta \tau_s^\delta w_{d,r,s}^\delta + 1 + e^{-2\alpha_{d,r,s}^\delta}})
    $$
    ,where $\bm{T}_{d\cdot,r,s}^\delta = \boldsymbol{\delta}_{1\cdot,r,s}\circ\cdots\circ \boldsymbol{\delta}_{d-1\cdot,r,s}\circ\boldsymbol{\delta}_{d+1\cdot,r,s}\circ\cdots\circ\boldsymbol{\delta}_{D\cdot,r,s}$, $m_{d\cdot,r,s,j}^\delta = \frac{1}{\sigma_e^2} \sum_{n=1}^{N_{\text{train}}} z_{ns}^2 \sum_{v'\in \mathcal{V}_{dj}} ( \bm{T}_{d\cdot,r,s}^\delta(v'))^2$, 
    and $n_{d\cdot,r,s,j}^\delta = \frac{1}{\sigma_e^2}\sum_{n=1}^{N_{\text{train}}} \langle \mathcal{Y}_{n,r,s,\cdot,dj}^{\delta}, \bm{T}_{d\cdot,r,s}^\delta \rangle$. For $j \in \{1, p_d\}$, the posterior distributions are expressed as: 
    \[
    \pi(\delta_{d\cdot,r,s,j}|-) = \mathcal{N}(\frac{n_{d\cdot,r,s,j}^\delta \tau_s^\delta w_{d,r,s}^\delta + e^{-\alpha_{d,r,s}^\delta}\delta_{d\cdot,r,s,j+1}}{m_{d\cdot,r,s,j}^\delta \tau_s^\delta w_{d,r,s}^\delta + 1}, \frac{\tau_s^\delta w_{d,r,s}^\delta }{m_{d\cdot,r,s,j}^\delta \tau_s^\delta w_{d,r,s}^\delta + 1}),
    \quad \text{for } j=1,
    \]
    \[
    \pi(\delta_{d\cdot,rs,j}|-) = \mathcal{N}(\frac{n_{d\cdot,r,s,j}^\delta \tau_s^\delta w_{d,r,s}^\delta + e^{-\alpha_{d,r,s}^\delta}\delta_{d\cdot,r,s,j-1}}{m_{d\cdot,r,s,j}^\delta \tau_s^\delta w_{d,r,s}^\delta + 1}, \frac{\tau_s^\delta w_{d,r,s}^\delta }{m_{d\cdot,r,s,j}^\delta \tau_s^\delta w_{d,r,s}^\delta + 1}),
    \quad \text{for } j=p_d.
    \]
    \item For a given dimension $d \in [1, D]$ and rank $r \in [1, R]$, the diagonal entries of the tensor margin covariance matrix $\mathbf{W}_{d,r,s}^\delta$, denoted as $w_{d,r,s}^\delta$, follow the generalized inverse-Gamma (gIG) posterior
    $$
    \pi(w_{d,r,s}^\delta \mid -) = \text{gIG} \left( \mu = 1 - \frac{p_d}{2}, \, \chi = \frac{c_{d,r,s}^\delta}{1 - e^{-2\alpha_{d,r,s}^\delta}}, \, \psi = \lambda_{d,r,s}^\delta \right),
    $$
    where 
    {\tiny $
    c_{d,r,s}^\delta = \sum_{s=1}^S \frac{1}{\tau_s^\delta} \left\{ \sum_{j=2}^{p_d-1} \left( 1 + e^{-2\alpha_{d,r}^\delta} \right) \| \delta_{d\cdot, rs, j} \|_2^2 + \| \delta_{d\cdot, rs, 1} \|_2^2 + \| \delta_{d\cdot, rs, p_d} \|_2^2 - 2e^{-\alpha_{d,r}^\delta} \sum_{j=1}^{p_d-1} \delta_{d\cdot, rs, j}^\top \ \delta_{d\cdot, rs, j+1} \right\}.
    $}
    \item The rate parameter $\lambda_{d,r,s}^\delta$ follows the conditional posterior {\small$\pi(\lambda_{d,r,s}^\delta \mid -) = \text{Ga} \left( a_{\lambda,s} + p_d, \, b_\lambda + \frac{p_d w_{d,r,s}^\delta}{2} \right).$}
    \item The global variance scale parameter $\tau_s^\delta$ follows the conditional posterior 
    $$
    \pi(\tau_s^\delta \mid -)
    = \text{gIG}\!\left(\mu = a_{\tau} - \frac{R(p_1 + \cdots + p_D)}{2},\, \chi = \sum_{r=1}^R \sum_{d=1}^D \boldsymbol{\delta}_{d\cdot,r,s}^\top (\mathbf{W}_{d,r,s}^\delta)^{-1} \boldsymbol{\delta}_{d\cdot,rs},\, \psi = 2b_\tau \right).
    $$
    \item The conditional posterior for parameter $\alpha_{d,r,s}^\delta$ satisfies the following
    $$
    \pi(\alpha_{d,r,s}^\delta \mid -) \propto (\alpha_{d,r,s}^\delta)^{a_\alpha - 1} (1 - e^{-2\alpha_{d,r,s}^\delta})^{-\frac{1}{2}S(p_d-1)} \exp \left[ -\frac{1}{2} \left( \boldsymbol{\delta}_{d\cdot, r,s}^\top (\mathbf{W}_{d,r,s}^\delta)^{-1} \boldsymbol{\delta}_{d\cdot, r,s} + 2b_\alpha \alpha_{d,r,s}^\delta \right) \right].
    $$
    Because the conditional posterior for $\alpha_{d,r,s}^\delta$ does not correspond to a closed-form distribution, $\alpha_{d,r,s}^\delta$ is sampled using a MH step as in step 5.

    \item Let $\mathcal{Y}_{n,r}^{\theta} = \mathcal{Y}_{n} - \widehat{\Gamma} - \widehat{\Theta}_r \odot \widehat{\mathcal{M}_{n,\cdot}(\mathcal{X}_{\mathcal{P},n})} - \sum_{s=1}^S \widehat{\mathcal{D}_s} z_{ns}$ be the rank $r^*$ specific residual corresponding to the $\Theta$ term, where $\widehat{\Theta}_r = \sum_{r' = 1, r' \neq r}^R \hat{\boldsymbol{\theta}}_{1\cdot,r'} \circ \cdots \circ \hat{\boldsymbol{\theta}}_{D\cdot,r'}$ where $\hat{\boldsymbol{\theta}}_{1\cdot,r'}, \cdots , \hat{\boldsymbol{\theta}}_{D\cdot,r'}$ are sampled from the most recent iteration, and $\widehat{\Gamma}$, $\widehat{\mathcal{D}_s}$ are taken from the most recently sampled instances of its tensor margins. The $j$-th element for tensor margin $\boldsymbol{\theta}_{d\cdot, r}$ for $j \in \{2,\cdots, p_d-1\}$ denoted $\theta_{d\cdot, r,j}$ follows the conditional posterior:
    $$
    \pi(\theta_{d\cdot,r,j}|-) = \mathcal{N}(\frac{n_{d\cdot,r,j}^\theta \tau^\theta w_{d,r}^\theta + e^{-\alpha_{d,r}^\theta}(\theta_{d\cdot,r,j-1}+\theta_{d\cdot,r,j+1})}{m_{d\cdot,r,j}^\theta \tau^\theta w_{d,r}^\theta + 1 + e^{-2\alpha_{d,r}^\theta}}, \frac{\tau^\theta w_{d,r}^\theta }{m_{d\cdot,r,j}^\theta \tau^\theta w_{d,r}^\theta + 1 + e^{-2\alpha_{d,r}^\theta}})
    $$
    ,where $\bm{T}_{d\cdot,r}^\theta = \boldsymbol{\theta}_{1\cdot,r}\circ\cdots\circ \boldsymbol{\theta}_{d-1\cdot,r}\circ\boldsymbol{\theta}_{d+1\cdot,r}\circ\cdots\circ\boldsymbol{\theta}_{D\cdot,r}$, \\
    $m_{d\cdot,r,j}^\theta = \frac{1}{\sigma_e^2} \sum_{n=1}^{N_{\text{train}}}  \sum_{v'\in \mathcal{V}_{dj}} (\widehat{\mathcal{M}_{n,\cdot}(\mathcal{X}_{\mathcal{P},n})}(v') \cdot\bm{T}_{d\cdot,r}^\theta(v'))^2$, \\
    and $n_{d\cdot,r,j}^\theta = \frac{1}{\sigma_e^2} \sum_{n=1}^{N_{\text{train}}} \sum_{v'\in \mathcal{V}_{n,dj}} \langle\mathcal{Y}_{n,r,\cdot,dj}^{\theta}, \bm{T}_{d\cdot,r}^\theta \rangle (v')$. For $j \in \{1, p_d\}$, the posterior distributions are expressed as:
    \[
    \pi(\theta_{d\cdot,r,j}|-) = \mathcal{N}(\frac{n_{d\cdot,r,j}^\theta \tau^\theta w_{d,r}^\theta + e^{-\alpha_{d,r}^\theta}\theta_{d\cdot,r,j+1}}{m_{d\cdot,r,j}^\theta \tau^\theta w_{d,r}^\theta + 1}, \frac{\tau^\theta w_{d,r}^\theta }{m_{d\cdot,r,j}^\theta \tau^\theta w_{d,r}^\theta + 1}),
    \quad \text{for } j=1,
    \]
    \[
    \pi(\theta_{d\cdot,r,j}|-) = \mathcal{N}(\frac{n_{d\cdot,r,j}^\theta \tau^\theta w_{d,r}^\theta + e^{-\alpha_{d\cdot,r}^\theta}\theta_{d\cdot,r,j-1}}{m_{d\cdot,r,j}^\theta \tau^\theta w_{d,r}^\theta + 1}, \frac{\tau^\theta w_{d,r}^\theta }{m_{d\cdot,r,j}^\theta \tau^\theta w_{d,r}^\theta + 1}),
    \quad \text{for } j=p_d.
    \]
    \item For a given dimension $d \in [1, D]$ and rank $r \in [1, R]$, the diagonal entries of the tensor margin covariance matrix $\mathbf{W}_{d,r}^\theta$, denoted as $w_{d,r}^\theta$, follow the generalized inverse-Gamma (gIG) posterior
    $$
    \pi(w_{d,r}^\theta \mid -) = \text{gIG} \left( \mu = 1 - \frac{p_d}{2}, \, \chi = \frac{c_{d,r}^\theta}{1 - e^{-2\alpha_{d,r}^\theta}}, \, \psi = \lambda_{d,r}^\theta \right),
    $$
    where {\tiny$
    c_{d,r}^\theta = \frac{1}{\tau^\theta} \left\{ \sum_{j=2}^{p_d-1} \left( 1 + e^{-2\alpha_{d,r}^\theta} \right) \| \theta_{d\cdot, r, j} \|_2^2 + \| \theta_{d\cdot, r, 1} \|_2^2 + \| \theta_{d\cdot, r, p_d} \|_2^2 - 2e^{-\alpha_{d,r}^\theta} \sum_{j=1}^{p_d-1} \theta_{d\cdot, r, j}^\top \theta_{d\cdot, r, j+1} \right\}.
    $}
    \item The rate parameter $\lambda_{d,r}^\theta$ follows the conditional posterior \\$\pi(\lambda_{d,r}^\theta \mid -) = \text{Ga} \left( a_\lambda + p_d, \, b_\lambda + \frac{p_d w_{d,r}^\theta}{2} \right).$
    \item The global variance scale parameter $\tau^\theta$ follows the conditional posterior 
    $$
    \pi(\tau^\theta \mid -) = \text{gIG} \left( \mu = a_\tau - \frac{R(p_1 + \cdots + p_D)}{2}, \, \chi = \sum_{r=1}^R \sum_{d=1}^D \boldsymbol{\theta}_{d\cdot, r}^\top (\mathbf{W}_{d,r}^\theta)^{-1} \boldsymbol{\theta}_{d\cdot, r}, \, \psi = 2b_\tau \right).
    $$
    \item The conditional posterior for lengthscale parameter $\alpha_{d,r}^\theta$ satisfies the following
    $$
    \pi(\alpha_{d,r}^\theta \mid -) \propto (\alpha_{d,r}^\theta)^{a_\alpha - 1} (1 - e^{-2\alpha_{d,r}^\theta})^{-\frac{1}{2}(p_d-1)} \exp \left[ -\frac{1}{2} \left( \boldsymbol{\theta}_{d\cdot, r}^\top (\mathbf{W}_{d,r}^\theta)^{-1} \boldsymbol{\theta}_{d\cdot, r} + 2b_\alpha \alpha_{d,r}^\theta \right) \right].
    $$
    Because the conditional posterior for $\alpha_{d,r}^\theta$ does not correspond to a closed-form distribution, $\alpha_{d,r}^\theta$ is sampled using a MH step as in step 5.
    
    \item The variance $\phi_1$ in the kernel matrices $\mathbf{K}_v$ follow the generalized inverse-Gamma) posterior:
    $\pi(\phi_{1}) = \text{Inv-Ga}(a_{\phi_1} + \frac{N \cdot|\mathcal{V}|}{2}, b_{\phi_1} + \sum_{v \in \mathcal{V}}\frac{1}{2} \bm{\mu_{v}}^\top \mathbf{K}_{v}^{-1} \bm{\mu_{v}})$, where $|\mathcal{V}|$ denotes the number of voxels within the voxel space $\mathcal{V}$ for each image, and $\bm{\mu_{n}}$ denotes $\widehat{\mathcal{M}_{\cdot,v}(\mathcal{X}_{\mathcal{P}}(v))}$ sampled from the most recent GP.
    \item The length-scale parameter $\phi_2$ in kernel matrices $\mathbf{K}_v$ has no closed-form distribution, $\phi_2$ is sampled using a MH step.
    \item The $\mathcal{M}_{\cdot,v}(\mathcal{X}_{\mathcal{P}}(v))$ for $v$-th voxel is sampled from a normal posterior distribution:
    $$ 
    \pi(\mathcal{M}_{\cdot,v}(\mathcal{X}_{\mathcal{P}}(v))| \mathcal{Y}_{n,r}^{\theta}) = N(\widehat{\Theta(v)}\cdot(\widehat{\Theta(v)}^2 \Sigma^{-1}+ \mathbf{K}_{v}^{-1})^{-1} \Sigma^{-1} \mathcal{Y}_{n,r}^{\theta},\quad (\widehat{\Theta(v)}^2 \Sigma^{-1}  + \mathbf{K}_{v}^{-1})^{-1}),
    $$
    where $\widehat{\Theta}$ is taken from the most recently sampled instances of its tensor margins, and $\widehat{\Theta(v)}$ is the tensor coefficient of $\widehat{\Theta}$ for the $v$-th voxel.
    \item Lastly, the residual variance term is updated as follows. Let
    $
    \mathcal{R}_{n} = \mathcal{Y}_{n} - \widehat{\Gamma} - \widehat{\Theta} \odot \widehat{\mathcal{M}_{n,\cdot}(\mathcal{X}_{\mathcal{P},n})} - \sum_{s=1}^S \widehat{\mathcal{D}_s}  z_{ns}
    $
    be the residual at the current iteration. Where $\widehat{\Gamma}$, $\widehat{\Theta}$,  $\widehat{\mathcal{D}_s}$ are taken from the most recently sampled instances of its tensor margins and  $\widehat{\mathcal{M}_{n,\cdot}(\mathcal{X}_{\mathcal{P},n})}$ is $\mathcal{M}_{n,\cdot}(\mathcal{X}_{\mathcal{P},n})$ sampled from the most recent GP based on the updated $\phi_1$ and $\phi_2$ from current iteration. Then the conditional posterior for noise variance parameter $\sigma_e^2$ for all voxels across all subjects is given by inverse-gamma distribution:
    $$
    \pi(\sigma_e^2|\text{-}) = \text{Inv-Ga} \left( a_e + \frac{N\cdot|\mathcal{V}|}{2}, \; b_e + \frac{1}{2} \sum_{n \in [1,N_{\text{train}}]} \sum_{v \in \mathcal{V}_n} \mathcal{R}_{n}^2(v) \right)
    $$
\end{enumerate}

\noindent\textbf{Modification due to incorporating individual-level masks: }In the Section \ref{sec:ADNI}, a individual level mask $\mathcal{S}_1$ was applied on individual images. Which lead to a minor changes in the posterior computation. In step 1, 6, 11, $v'\in \mathcal{V}_{dj}$ should be replaced by $v'\in \mathcal{V}_{n,dj}$, where $v'\in \mathcal{V}_{n,dj}$ denotes the voxel space for subject $n$. And step 16, 19  $\mathcal{V}$ should be replaced by $\sum_{n=1}^N|\mathcal{V}_n|$, where $|\mathcal{V}_n|$ denotes the number of voxels within mask $\mathcal{S}_1$ of each subject.

\subsection{Additional Results}
\label{s:addi}
\begin{figure}
    \centering
    \includegraphics[width=0.8\linewidth]{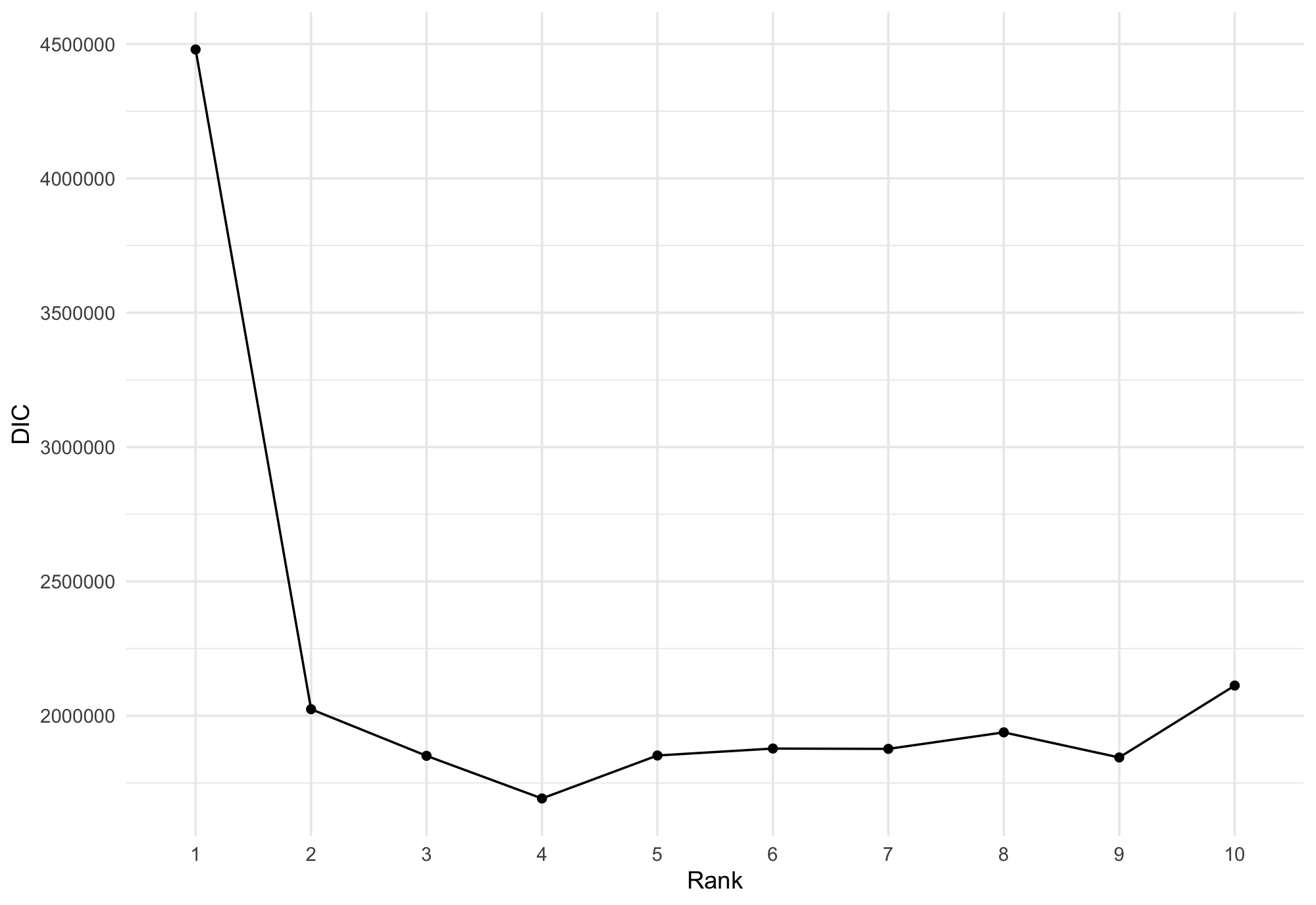}
    \caption{DIC values for different tensor rank choices for simulation setting 3.a.ii}
    \label{fig:dic}
\end{figure}

\begin{figure}
    \centering
    \includegraphics[width=0.8\linewidth]{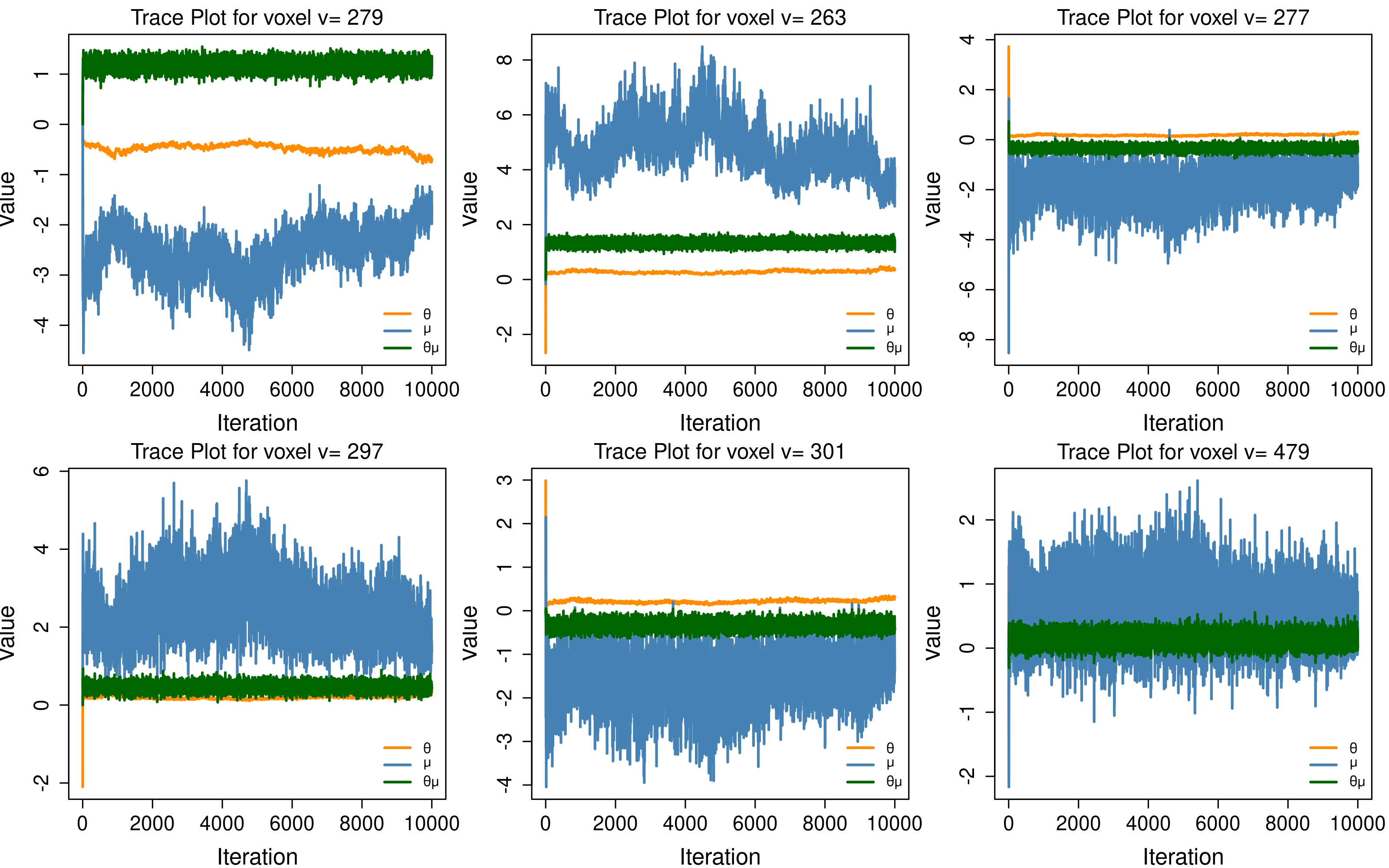}
    \caption{Example traceplots for parameters in simulation. The blue curve corresponds to $\Theta(v)$ ($\theta$), the orange curve corresponds to $\mathcal{M}_{n,v}(\mathcal{X}_{\mathcal{P},n}(v))$ ($\mu$), and the green curve corresponds to $\Theta(v) \cdot \mathcal{M}_{n,v}(\mathcal{X}_{\mathcal{P},n}(v))$ ($\theta\mu$) for the given voxels.}
    \label{fig:traceplot}
\end{figure}

We present additional simulation results omitted from the main text. Figure~\ref{fig:dic} illustrates DIC values across candidate tensor ranks R, with the optimal rank chosen by minimizing the DIC. Figure~\ref{fig:traceplot} shows MCMC trace plots for $\Theta(v)$, $\mathcal{M}_{n,v}(\mathcal{X}_{\mathcal{P},n}(v))$ and $\Theta(v) \cdot \mathcal{M}_{n,v}(\mathcal{X}_{\mathcal{P},n}(v))$, demonstrating stable posterior sampling.

\begin{table}[p]
\centering
\caption{MCMC diagnostics for the BTOT-VC model based on Geweke $z$-scores. Results are presented as the mean (SD) of median Geweke $z$-scores calculated across all voxels. Specifically, diagnostics were computed for the terms $\Theta(v)\cdot \mathcal{M}(\mathcal{X}_{n,\text{patch}}(v))$ (Scenario 1), $\Theta(v)\cdot \mathcal{M}(\mathcal{X}_n(v))$ (Scenario 2), $\Theta(v)\cdot \sin\!\left(\sum_{v' \in \text{patch}} \mathcal{X}_{n,v'}(v)\right)$ (Scenario 3), and $\Theta(v)\cdot \mathcal{X}_n(v)$ (Scenario 4). All absolute Geweke $z$-scores were below 1.96, indicating good convergence of the MCMC algorithm.}
\label{tab:geweke}
\footnotesize
\begin{tabular}{lc}
\toprule
\textbf{Scenario} & \textbf{Geweke $z$ (Mean (SD))} \\
\midrule
1.a.i  & 0.150 (0.188) \\
1.a.ii & 0.051 (0.016) \\
1.b.i  & -0.009 (0.082) \\
1.b.ii & -0.034 (0.054) \\
1.c.i  & 0.076 (0.112) \\
1.c.ii & 0.072 (0.061) \\
1.d.i  & -0.053 (0.059) \\
1.d.ii & 0.062 (0.022) \\

\midrule
2.a.i  & 0.020 (0.101) \\
2.a.ii & 0.137 (0.220) \\
2.b.i  & 0.020 (0.101) \\
2.b.ii & 0.099 (0.059) \\
2.c.i  & 0.021 (0.066) \\
2.c.ii & 0.123 (0.203) \\
2.d.i  & -0.024 (0.151) \\
2.d.ii & 0.093 (0.074) \\

\midrule
3.a.i  & 0.023 (0.117) \\
3.a.ii & 0.054 (0.013) \\
3.b.i  & 0.019 (0.089) \\
3.b.ii & 0.072 (0.017) \\
3.c.i  & -0.058 (0.122) \\
3.c.ii & 0.056 (0.015) \\
3.d.i  & 0.025 (0.104) \\
3.d.ii & 0.066 (0.029) \\

\midrule
4.a.i  & 0.027 (0.165) \\
4.a.ii & 0.036 (0.028) \\
4.b.i  & 0.093 (0.078) \\
4.b.ii & 0.024 (0.086) \\
4.c.i  & -0.065 (0.217) \\
4.c.ii & 0.074 (0.061) \\
4.d.i  & 0.045 (0.361) \\
4.d.ii & 0.016 (0.058) \\
\bottomrule
\end{tabular}
\end{table}

\end{document}